\documentclass[pra,showpacs,twocolumn,superscriptaddress,floatfix,aps]{revtex4-2}
\usepackage{amsmath,amssymb,bm}
\usepackage{graphicx}
\usepackage{physics}
\usepackage[T1]{fontenc}
\usepackage{comment}
\usepackage{float}
\usepackage{mathtools}
\usepackage[normalem]{ulem}
\usepackage{hyperref}
\usepackage{soul}
\usepackage{braket}
\usepackage[utf8]{inputenc}
\hypersetup{colorlinks,%,%
 linkcolor=blue,%
 citecolor=blue,%
 urlcolor=blue}
\usepackage{color}
\usepackage{xcolor}
\usepackage{soul,xcolor} %Strike through this text

\def\dar{\downarrow}
\def\upar{\uparrow}

\begin{document}
\setstcolor{red}

%\title{Controlling soliton dynamics through inhomogeneous Rabi-field in a spin-$1/2$ condensate }
\title{Rabi-Engineered spin-selective transport of bright soliton in a binary spin-1/2 condensate}
%\title{Soliton Spintronics through inhomogeneous Rabi-coupled Bose–Einstein Condensate}

\author{Swarup K. Sarkar}
\email{swarup69190@gmail.com}
%\affiliation{Department of Physics, Indian Institute of Technology Guwahati, Guwahati 781039, Assam, India}
\affiliation{Centre for computational and data sciences, Indian Institute of Technology Kharagpur, Kharagpur, West Bengal 721302, India}

\author{Sh. Mardonov}
\affiliation{New Uzbekistan University, Movarounnahr str. 1, Tashkent 100000, Uzbekistan}
\affiliation{Ulugh Beg Astronomical Institute, Tashkent, 100052, Uzbekistan}

\author{E. Ya. Sherman}
\affiliation{Department of Physical Chemistry, University of the Basque Country UPV/EHU, 48940 Leioa, Spain}
\affiliation{IKERBASQUE, Basque Foundation for Science, Bilbao, Spain}
\affiliation{EHU Quantum Center, University of the Basque Country EHU, 48940 Leioa, Spain}
% \email{Corresponding author, evgeny.sherman@ehu.eus}
%\footnote{Corresponding author}

\author{Pankaj K. Mishra}
%\email{pankaj.mishra@iitg.ac.in}
\affiliation{Department of Physics, Indian Institute of Technology Guwahati, Guwahati 781039, Assam, India}
\date{\today}

% \date{\today}

\begin{abstract}
Spintronics is usually concentrated on spin-dependent dynamics of single particles such as electrons or holes in solids. Extending this concept to nonlinear matter waves offers an interesting possibility of controlling the transport of collective, particle-like excitations through their spin degrees of freedom.  Recent advances in Bose-Einstein condensates have provided a versatile platform for realizing nonlinear spin-1/2 systems that support particle-like objects such as bright solitons, making studies of the motion of these solitons under external fields an interesting and important problem. Here, we study the dynamics of bright solitons with different spin symmetries moving in a coordinate-dependent Rabi field, which acts analogously to a spatially varying magnetic field.  In that setup, the velocity perturbed soliton motion possesses very interesting scattering dynamics for both perfect and lifted Manakov symmetry. Lastly, we compare both the velocity-controlled scattering problem with the well-known Rabi quenching protocol. This comparison provides a powerful approach to control the soliton dynamics that may open an avenue to control nonlinear spintronics with solitons as spin carriers. 
\end{abstract}

\date{\today}

\maketitle

\section{Introduction}
Spintronics, the spin-dependent transport of charge carriers, is one of the most rapidly developing fields in modern fundamental and applied physics~\cite{Zutic:2004}. The usual object of study in spintronics is spin-dependent dynamics of electrons and holes in metal-based or semiconducting structures. Motion of charge carriers in position-dependent magnetization regions such as, e.g., domain walls, over which a ferromagnetic order parameter smoothly reverses, strongly affects their transport, producing measurable domain-wall resistance and related magnetoresistance in solid-state systems. These effects are  extensively studied experimentally  \cite{Gregg:1996,Rudiger:1999,Danneau:2002} and analyzed theoretically~\cite{Dugaev:2006}, with one of the important problems being a controllable spin-dependent delivery of carriers. Possibly strong nonlinearities in these systems produce a wide variety of magnetic solitons \cite{Kosevich:1990}. 

These fundamental concepts of spin transport are not restricted solely to solids. The experimental realization of spin-orbit coupled Bose--Einstein condensates (BECs) has established them as a versatile platform for investigating a broad range of spin-dependent phenomena~\cite{cheng:2014,Li:2016,Sarkar:2024,Xi:2015,Zhang:2022,Sarkar:2025} enabling direct exploration of solid-state spintronics analogues \cite{Dalibard:2011,lin:2011,galitski:2013,Goldman:2014}. Owing to their high controllability, these systems provide a unique setting to explore spin transport and manipulation. Unlike single-particle systems, the pseudospin-$1/2$ binary condensates exhibit nonlinear spin-dependent interactions together with coherent intercomponent couplings, such as Rabi coupling, producing a linear connection between spin components, giving rise to a wealth of collective phenomena. For example, uniform Rabi coupling breaks the phase symmetry between the two components, enabling the spontaneous formation of relative-phase domain walls and other topological excitations~\cite{sushanto:2008,Ihara:2019}. It enables population exchange in optical lattices \cite{Trombettoni:2009}, vectorizes dark solitons \cite{Liu:2012}, and affects miscibility of spin components  \cite{Merhasin:2005, Merhasin:2005phy_script,Gligori:2010, Chiquillo:2018}.  Also, it can produce rogue waves in BECs~\cite{Kanna:2019}, stabilize soliton-like states under time-modulated coupling ~\cite{sushanto:2008}, and inhibit rotation-induced vortex formation ~\cite{Zhao:2021rot_bec}. The lifted Manakov symmetry~\cite{Cheng_spat:2010} can produce spatially asymmetric localization of the condensate~\cite{Sarkar:2025b}. Certain topological structures can be realized in various setups with corresponding parameters  \cite{Son:2002,Qu:2017}. 

The physics of systems with spatially engineered  Rabi coupling is even more intriguing. Within the cold-atom framework, spatially engineered Rabi coupling has been investigated in several contexts beyond conventional Rabi coupling. Periodic inhomogeneity of the Rabi coupling, the ``Rabi lattice'', was shown to produce band-gap structures supporting families of gap solitons~\cite{Chen:2017}, while localized Rabi coupling acts as an effective attractive potential capable of trapping and manipulating solitons~\cite{Mardonov:2014}.  Recently, it was demonstrated that magnetic domain walls in spinor BECs can act as efficient scattering centers for elementary excitations, exhibiting controllable propagation dependent on the structure of the domain wall~\cite{Zhao:2026}.

Since Rabi fields strongly influence the properties and dynamics of multicomponent Bose--Einstein condensates, a natural question arises: how does a soliton propagate and scatter in the presence of a spatially nonuniform Rabi field? In this work, we address this problem by investigating the dynamics of a soliton traversing a coordinate-dependent Rabi coupling, analogous to a conventional domain wall in solid-state systems. As we shall demonstrate, even this minimal setting gives rise to remarkably rich dynamics in both the density and spin sectors, with the scattering behavior being strongly governed by the symmetry of the intercomponent interactions.

Although soliton scattering from external potentials has been comprehensively studied in ultracold gases (see, e.g., \cite{Otajonov:2026}), these studies almost exclusively consider spin-independent potentials that couple identically to both pseudospin components. Consequently, they do not capture the spin-selective transport induced by spatially varying intercomponent coupling, nor do they reveal the qualitative role played by interaction symmetry. In contrast, the present work focuses on spin-dependent scattering processes, with particular emphasis on the distinct transport characteristics emerging in the Manakov and lifted Manakov symmetry regimes.

This paper is organized as follows. In Sec.~\ref{sec:2}, we formulate the mean-field model describing the soliton propagation with two-component Gross-Pitaevskii equations and present the main observables to characterize the system of interest. Section~\ref{sec:3} presents the numerical result related to scattering process in general due to the effect of Rabi barriers in detail in terms of preserved and lifted Manakov symmetry of nonlinearities. Finally, we compare the dynamics originated from velocity perturbation and Rabi field quenching for different Rabi coupling. Section~\ref{sec:4} presents the summary and conclusion.

\begin{figure}
    \centering
    \includegraphics[width=\linewidth]{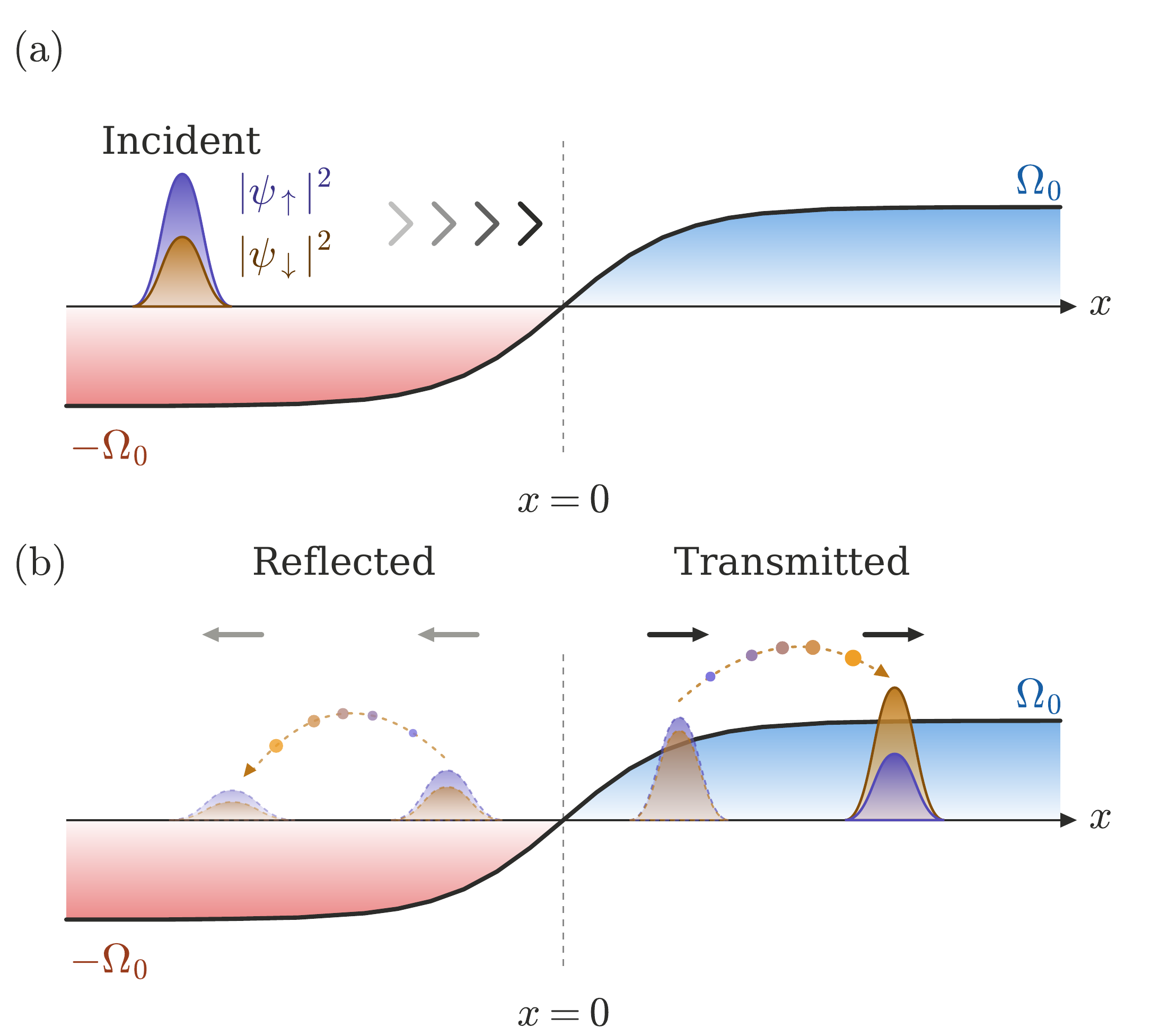}
    \caption{Schematic illustration of the bright soliton crossing a spatially inhomogeneous Rabi field. (a) A two-component incident soliton with densities $|\psi_{\upar}|^{2}$ and $|\psi_{\dar}|^{2}$ approaching the region with position-dependent Rabi coupling.  (b) Transmitted and reflected solitons with the time- and position-dependent population transfer between $\dar$ and $\upar$ spin components.  Figure is for illustration purpose only and not to scale.}
    \label{fig:illustration}
\end{figure}
%%%%%%%%%%%%%%%%%%%%%%%%%%%%%%%%%%%%%
\section{Mean-field model and  observables}
\label{sec:2}
To investigate Rabi-engineered spin-selective transport, we first formulate the mean-field description of a binary spin-1/2 condensate subject to a spatially varying coherent coupling. The Rabi field provides the key control parameter of the transport process: its magnitude determines the strength of the coupling between the two spin components, while its spatial variation creates an effective interface encountered by the propagating soliton. We therefore introduce the coupled Gross-Pitaevskii equations and the associated energy contributions and the observables that separately characterize spatial transport along with internal spin dynamics.

\subsection{Coupled Gross-Pitaevskii equations}
We consider a cigar-shaped pseudospin-$1/2$ condensate strongly trapped in the transverse directions and located in the presence of an inhomogeneous Rabi field along the $x-$ axis. The condensate can be modeled using two-component Gross-Pitaevskii equations coupled through a spatially dependent Rabi coupling as follows~\cite{Santos:2021,Chen:2017,Sarkar:2025}: 
\begin{subequations}%
\label{eqn1}%
\begin{align}%
% \begin{split}
 {\mathrm i} \frac{\partial \psi_{\upar}}{\partial t} = & \left[-\frac{1}{2}\frac{\partial^{2}}{\partial x^{2}} + 
 g_{\upar \upar} \vert \psi_{\upar} \vert ^{2} + 
 g_{\upar\dar} \vert \psi_{\dar} \vert ^{2}\right]\psi_{\upar}  + 
 \Omega(x) \psi_{\dar}
  , \label{eqn1(a)}\\ 
 {\mathrm i} \frac{\partial \psi_{\dar}}{\partial t} = & \left[-\frac{1}{2}\frac{\partial^{2}}{\partial x^{2}} + 
 g_{\dar\dar} \vert \psi_{\dar} \vert ^{2} + g_{\dar \upar} \vert \psi_{\upar}\vert^{2}\right]\psi_{\dar} + 
 \Omega(x) \psi_{\upar}, \label{eqn1(b)} 
\end{align}
\end{subequations}%
where $\psi_{\upar}$ and $\psi_{\dar}$ $(\psi_{\upar,\dar}\equiv \psi_{\upar,\dar}(x,t))$ represent the pseudo spin-up and spin-down components of the condensate wavefunction ${\bm\psi}=\left(\psi_{\upar},\psi_{\dar}\right)^{\rm T}$, respectively, where ${\rm T}$ stands for transposition. For stationary states $\psi_{\upar,\dar}(x,t)=\psi_{\upar,\dar}(x)\exp(-{\mathrm i}\mu t),$ where $\mu$ is the chemical potential. Here, $g_{\upar\upar}$ and $g_{\dar\dar}$ are the intra-species interaction, $g_{\upar\dar}$ and $g_{\dar\upar}$ are inter-species interactions, and $\Omega(x)$  is the position-dependent Rabi field. In the following, we omit the $(x,t)$ notation in  $\psi_{\upar (\dar)}$ to avoid further confusion. The ground state can be obtained using imaginary time propagation of Eq.~(\ref{eqn1(a)}-\ref{eqn1(b)}) with the normalization, $N_\upar + N_\dar = 1$, where $N_j = \int_{-\infty}^{\infty}|\psi_j|^{2}dx$ and $j \equiv \{\upar,\dar \}$. 

To obtain the dimensionless Eq.~(\ref{eqn1(a)})-(\ref{eqn1(b)}), we consider the transverse harmonic oscillator length $a_{\perp} = \sqrt{\hbar/(m\omega_{\perp})}$ as a characteristic length scale with $\omega_{\perp}$ as the transverse harmonic trapping frequency, $\omega_{\perp}^{-1}$ as the timescale and $\hbar \omega_{\perp}$ as the characteristic energy scale. The interaction parameters can be defined in terms of {$g_{{\upar\upar},({\dar\dar})}= 2\mathcal{N}a_{{\upar\upar},({\dar\dar})}/a_{\perp}$,} and $g_{\upar\dar} = 2\mathcal{N}a_{\upar\dar}/a_{\perp}$, where, $a_{{\upar\upar},({\dar\dar})}$ and $a_{\upar\dar}$ represent the intra- and inter-component scattering lengths, respectively, and $\mathcal{N}$ represents the total number of atoms in the condensate. The dimensionless Rabi coupling amplitude $\Omega_0$ is defined in the units of $2\omega_{\perp}$ with the wavefunction being rescaled with $\sqrt{a_{\perp}}.$

The total soliton energy is defined as the sum of kinetic, self-interaction, and Rabi energies $E=E_{\rm kin}+E_{\rm int}+E_{\Omega}$ defined as:
\begin{eqnarray}
   &&\hspace{-0.6cm} E_{\rm kin}=  \int_{-\infty}^{\infty} \frac{1}{2}\left(|\psi^{\prime}_\upar|^2 + |\psi^{\prime}_\dar|^2 \right)dx, \label{eq:Ekin}\\
     &&\hspace{-0.6cm} E_{\rm int} = \int_{-\infty}^{\infty} \left(\frac{g_{\upar \upar}}{2}|\psi_{\upar}|^4  + \frac{g_{\dar \dar}}{2}| \psi_{\dar}|^4 + g_{\upar\dar}|\psi_{\upar}\psi_{\dar}|^2\right)dx, \label{eq:Ese}\\
  &&\hspace{-0.6cm} E_{\Omega} =  \int_{-\infty}^{\infty} \Omega(x) \left( \psi^{*}_\dar \psi_\upar + \psi^{*}_\upar \psi_\dar\right)\, dx, \label{eq:ER}
\end{eqnarray}
with $\psi_j^\prime \equiv{\partial \psi_j}/{\partial x}.$ Here the interspecies interactions are same for both component i.e.,  $g_{\upar \dar } = g_{\dar \upar}$.
The inhomogeneous Rabi coupling is incorporated with barrier-shaped $\Omega(x)$ in the form:
\begin{equation} \label{eqn2}
    \Omega(x) = \Omega_0 \tanh(x).     
\end{equation}

The motivation for choosing a spatially varying Rabi coupling, $\Omega(x)$, extends beyond a mathematically convenient smooth interface. A sign-changing profile represents the simplest realization of a Rabi domain wall, where the coherent coupling between the two pseudospin components reverses smoothly across $x=0.$ Such an interface provides a minimal setting for exploring domain-wall-mediated spin transport in a nonlinear Bose-Einstein condensate, analogous to the role of magnetic domain walls in solids~\cite{Rudiger:1999, Danneau:2002, Dugaev:2006}.  Motivated by these analogies, we investigate the scattering of a propagating bright soliton by a Rabi interface. Unlike the conventional scattering by a potential barrier, the present Rabi barrier modifies the internal coupling between the condensate components. Furthermore, lifted Manakov symmetry among interactions makes the system nonintegrable and leads to a nontrivial soliton scattering dynamics due to a competition between kinetic energy, self-interaction, and the spatially varying coherent coupling. This setting enables us to examine how the Rabi domain wall governs soliton reflection, transmission, and spin conversion in a binary BEC.

The initial state has the form 
\begin{equation}
{\bm \psi}(x,0) = {\bm \psi}_{{\rm gs}}(x-x_{0})\exp(ivx),
\label{eq:psi_dyn}
\end{equation}
where $x_{0}<0$ with $|x_{0}|\gg\,1$  and ${\bm \psi}_{{\rm gs}}(x-x_{0})$ is the corresponding soliton ground state for the Rabi field $-\Omega_{0},$ and $v$ denotes the initial soliton velocity. Thus, the soliton acquires, in addition to contributions \eqref{eq:Ekin}-\eqref{eq:ER}, kinetic energy $E_k = mv^{2}/2.$ The time $t > 0$ demonstrates the subsequent real-time dynamics. Here, the generation of a solitonic ground state with $-\Omega_0$ associated with the negative constant value of Eq.\eqref{eqn2} in $x < 0$ can act as the stationary state in real-time propagation along with $\Omega(x)$.

The Rabi-coupled BEC of interest can be realized by using hyperfine states of $^{87}$Rb atoms as pseudo spin-up $\ket{\upar} \equiv \ket{F = 1, m_F = 0}$ and spin-down $\ket{\dar} \equiv \ket{F = 1, m_F = -1}$, which are coupled by a pair of Raman lasers. The recoil energy $E_{rc} = \hbar^{2}k_{p}^{2}/2m,$  where $k_{p}$ is the photon wavevector, a relevant scale for tuning the Rabi coupling ~\cite{lin:2011, Jim:2015}.  The intra- and inter-spin scattering lengths, typically of the order of 5 nm  can be further controlled by using Feshbach resonances.

The spatial profile of Rabi coupling can be engineered with the standard Raman-dressing scheme used to generate spin–orbit coupling in $^{87}$Rb condensates~\cite{lin:2011, Manakov:1974}. The intensity profile can be imprinted by shaping the Raman beams with programmable optics, for example, with spatial light modulators or digital micromirror devices which are routinely used to sculpt smooth, arbitrary optical potentials in ultracold atom experiments~\cite{Gauthier:2016,Gauthier:2021}. These beam-shaping techniques provide independent control over the amplitude and phase of the laser beam profile~\cite{Nogrette:2014,Goorden:2014}, which is precisely the capability required here, since the magnitude and sign of the Rabi coupling are determined by the intensity and relative phase of the Raman beams, respectively. A fixed $\pi$ phase offset has previously been used to realize a negative coupling in theoretical studies of SO-coupled binary systems~\cite{cheng:2014, Sarkar:2024}. More generally, a smoothly varying spatial phase difference between the Raman beams provides a natural route to engineering a continuously changing Rabi coupling. The width of this transition would be determined by the optical resolution, which is typically of the order of a few microns and is comparable to the transverse harmonic-oscillator length $a_{\perp}$ for typical trapping frequencies. Consequently, the corresponding dimensionless barrier width is expected to be of the order of unity for realistic experimental parameters. This provides a feasible route toward experimental realization, which could be explored in detail in future ultracold atom experiments.

\subsection{Definition of physical observables} \label{sec:IIB}
 The transport process involves two coupled degrees of freedom: the spatial partition of the soliton into reflected and transmitted components and the redistribution of atoms between the two-spin components. To distinguish these effects, we introduce observables that can resolve both spatial and spin degrees of freedom.
 
Since we are considering the reflection/transmission of the soliton in the presence of space-dependent Rabi coupling, we separate the entire $ x$-axis into two subspaces: $L$ subspace for $x<0$ and $R$ subspace for $x>0$, and define corresponding subspace-specific spin-projected observables. Before the soliton collision with the $\Omega(x)$ barrier, only the left subspace is populated, while after the collision both subspaces can be populated, with their relative occupancies determining the reflection and transmission probabilities.

We begin with definition of subspace- and spin-dependent atomic populations:
\begin{equation}
N_{Lj}=\int_{-\infty}^{0} \vert \psi_{j} \vert ^{2} dx, \quad N_{Rj}=\int_{0}^{\infty} \vert \psi_{j} \vert ^{2} dx
 \label{eqn:N}
\end{equation} 
with $j=\upar,\dar$ and $N_{L\upar}+N_{L\dar}+N_{R\upar}+N_{R\dar}=1,$ and corresponding quantities $N_{L}=N_{L\dar}+N_{L\upar}$,
$N_{R}=N_{R\dar}+N_{R\upar},$ $N_{\dar}=N_{L\dar}+N_{R\dar},$ and $N_{\upar}=N_{L\upar}+N_{R\upar}.$

To characterize the spin-projected profiles of the incident, scattered, and transmitted solitons, we further introduce corresponding inverse participation ratios $\chi_{Lj}$ and $\chi_{Rj}$ in the respective subspaces as:
\begin{align}
    \chi_{Lj}=\frac{1}{N_{Lj}^{2}}\int_{-\infty}^{0} \vert \psi_{j} \vert^{4} dx; \quad \chi_{Rj}=\frac{1}{N_{Rj}^{2}}\int_{0}^{\infty} \vert \psi_{j} \vert^{4} dx 
    \label{eq:ipr}
\end{align}
To complete the scattering picture, we also compute the spin observables 
\begin{align}
\label{eq:sigx}
\langle \sigma_{i} \rangle = 
\int_{-\infty}^{\infty}\bm{\psi}^{\dagger}\sigma_{i}{\bm \psi} dx,
\end{align}
with $\sigma_{i}$ ($i=x,y,z$) being the Pauli matrices, to characterize the spin polarization of the condensate. Here $\langle \sigma_x \rangle =1$ ($\langle \sigma_x \rangle =0$) corresponds to a fully polarized (unpolarized) by the Rabi coupling spin state.

\section{Numerical Results}
\label{sec:3}

Here, we present a detailed analysis of the ground state soliton and its velocity-dependent evolution including transmission and reflection as it traverses the inhomogeneous Rabi field, as illustrated in Fig.~\ref{fig:illustration}. We consider two distinct scenarios: the Manakov symmetry soliton and the Manakov symmetry lifted by spin-dependent nonlinearities. In both cases, we first obtain the soliton ground state centered at $x_{0}$ using imaginary-time propagation. Subsequently, a finite velocity is imprinted in the ground state, as described by Eq.~\eqref{eq:psi_dyn}, and its real-time evolution is studied as the soliton moves toward $x > 0,$ where it experiences the position-dependent Rabi field.       

\subsection{Calculation details}
To obtain the soliton ground state by using imaginary time propagation, we numerically solve the coupled GPEs (Eqs.~(\ref{eqn1(a)})-(\ref{eqn1(b)})) by applying split-step Fourier spectral method with space step $\Delta x = 0.025$, time step $\Delta t = 6.25\times 10^{-4},$ at an interval $-X_{\max}<x<X_{\max}$ with $X_{\max}=204.8.$ The initial condition is considered as a Gaussian profile centered at the initial soliton position $x_{0}.$ We study the subsequent condensate motion by real-time dynamics with the same space and time steps.

In the following, we first discuss the soliton with Manakov symmetry and then move to the cases where this symmetry is lifted through unequal spin-dependent interactions.

\subsection{Soliton with Manakov symmetry}

Manakov model, originally introduced in 1974 \cite{Manakov:1974} describes an integrable system in the case of multicomponent GPE's with equal intra- and inter-component interactions \cite{Menyuk:1988, Evangel:1992}. In spinor condensates, including those with spin-orbit coupling, the Manakov symmetry considerably simplifies studies of their dynamics by the fact that the self-interaction depends on the total density\cite{Tokatly2010}. 

To implement the Manakov symmetry we choose all interactions to be equal as $g=g_{\upar \upar} = g_{\dar \dar} = g_{\upar \dar} = g_{\dar \upar}.$ Here the ground state spin-densities are the same for both components irrespective of $\Omega_{0}$ and $g.$ Here $\psi_{{\rm gs},\upar}(y)=\psi_{{\rm gs},\dar}(y)=\sqrt{|g|}/2\sqrt{2}{\rm cosh}(y/2g)$  with $y=x-x_{0},$ $\langle \sigma_{x}\rangle = 1,$ and total energy $E= -\Omega_{0}-g^{2}/24.$ 

To numerically study the soliton motion, we first generate the ground state $\psi_{{\rm gs}}(x)$ at $x_{0} = -30$ with $\Omega_{0}=0.05.$  It is followed by the implementation of velocity to the ground state as described in Eq.~\eqref{eq:psi_dyn} along with $\Omega(x).$. The spatially inhomogeneous Rabi field for the moving soliton acts as an effective scattering potential dependent on the soliton's spin state.  The Manakov case possesses an exact SU(2) symmetry of interactions, resulting in a sole dependency of self-interaction energy on the total density. Consequently, the soliton maintains $\langle\sigma_{x}\rangle=1$ during propagation and upon approaching the Rabi barrier, where it can encounter a significant reflection.
\begin{figure}[!htp]
    \centering
    \includegraphics[width=\linewidth]{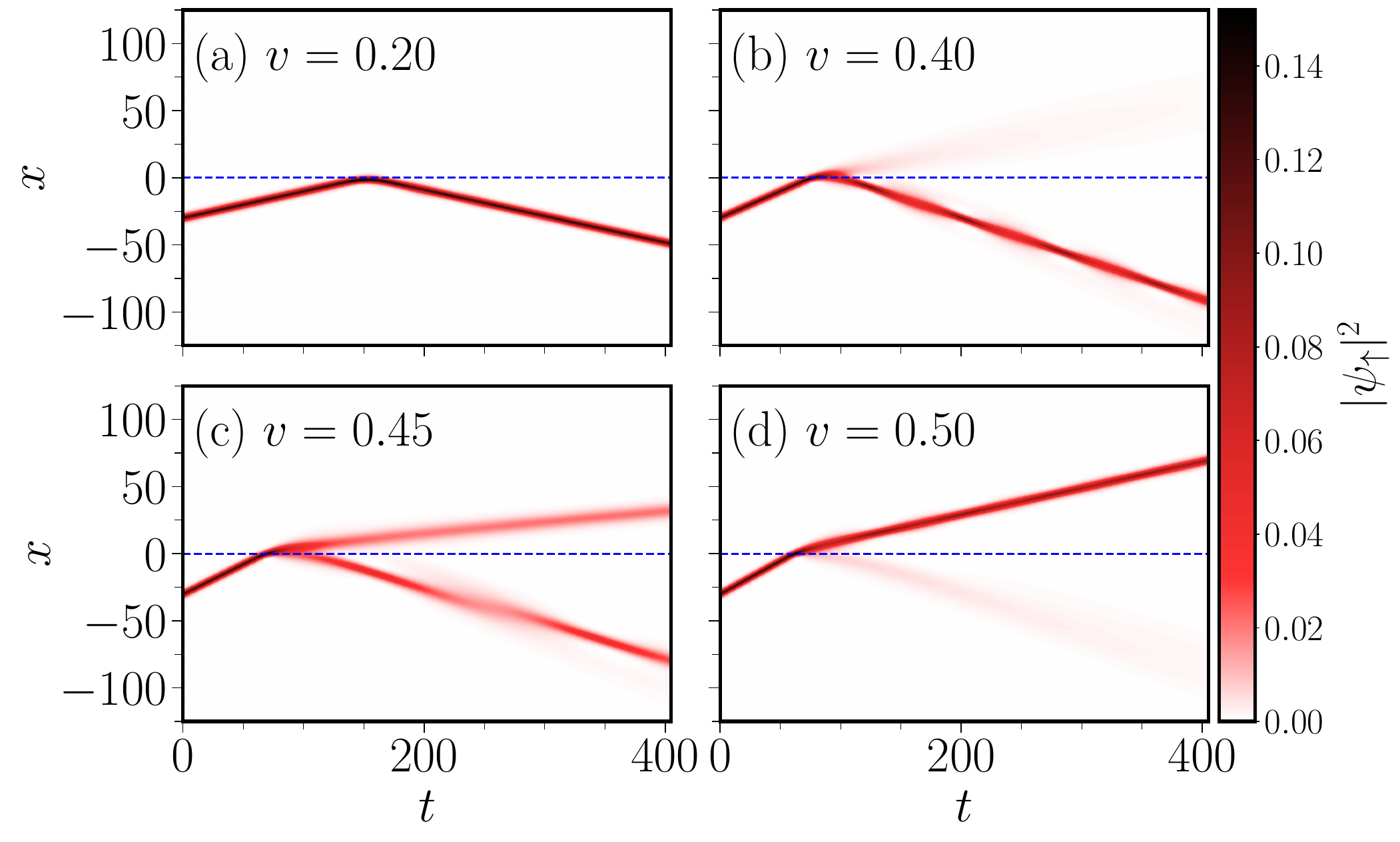}
    \caption{Pseudo-color-map representation of condensate density $|\psi_{\upar}|^2$ in $(t,x)$ plane for different velocities: (a)$v = 0.2$, (b) $v = 0.4$, (c) $v = 0.45$, and (d) $v = 0.5$. For velocity $v \leq 0.4$, the soliton mostly gets reflected. For $v \geq 0.45$, the soliton gets partially reflected and partially transmitted. For larger velocities, the transmission probability increases. Other parameters are $g_{\upar \upar} = g_{\dar \dar} = g_{\upar \dar} = g_{\dar \upar} = -1.0, \Omega_0 =0.05,$ and initial $x_0 = -30.$ }
    \label{fig:den-mnkv}
\end{figure}

Figure~\ref{fig:den-mnkv} shows the evolution of the condensate density $|\psi_{\upar}|^2$ in the $(t,x)$ plane associated to different velocity perturbations. Since the Manakov symmetry leads to  $|\psi_{\upar}|^2=|\psi_{\dar}|^2$ equality, we only show the spin-$\upar$ component. For $v = 0.2$ in \ref{fig:den-mnkv}(a), and $0.40$ in \ref{fig:den-mnkv}(b), the soliton mostly gets reflected after hitting the Rabi barrier at $\tau \approx |x_0|/ v \approx 150,$ and $ 75.0,$ respectively with $x_0 = -30$ being the initial soliton position. The density evolution at $v = 0.45$ demonstrates that $\sim 50\%$ of the atoms from the soliton is reflected, and the other half gets transmitted through hitting the barrier at $\tau \approx  66.7.$ Also, the reflected counterpart at $t > \tau$ propagates faster than the transmitted one. It is because of the reflected part remain in the same polarization state as the initial value, whereas the transmitted part follows a change in polarization Rabi field changes. 

 As we increase the velocity to $v = 0.5$ in Fig.~\ref{fig:den-mnkv}(d), the transmission probability increases while the population of the reflected part decreases accordingly. This effect can be seen with the analogy to quantum mechanics, where the velocity of the reflected wave is $-v$ while the velocity of the transmitted wave is $\sqrt{v^{2}-4\Omega_{0}}.$  
\begin{figure}
    \centering
    \includegraphics[width=\linewidth]{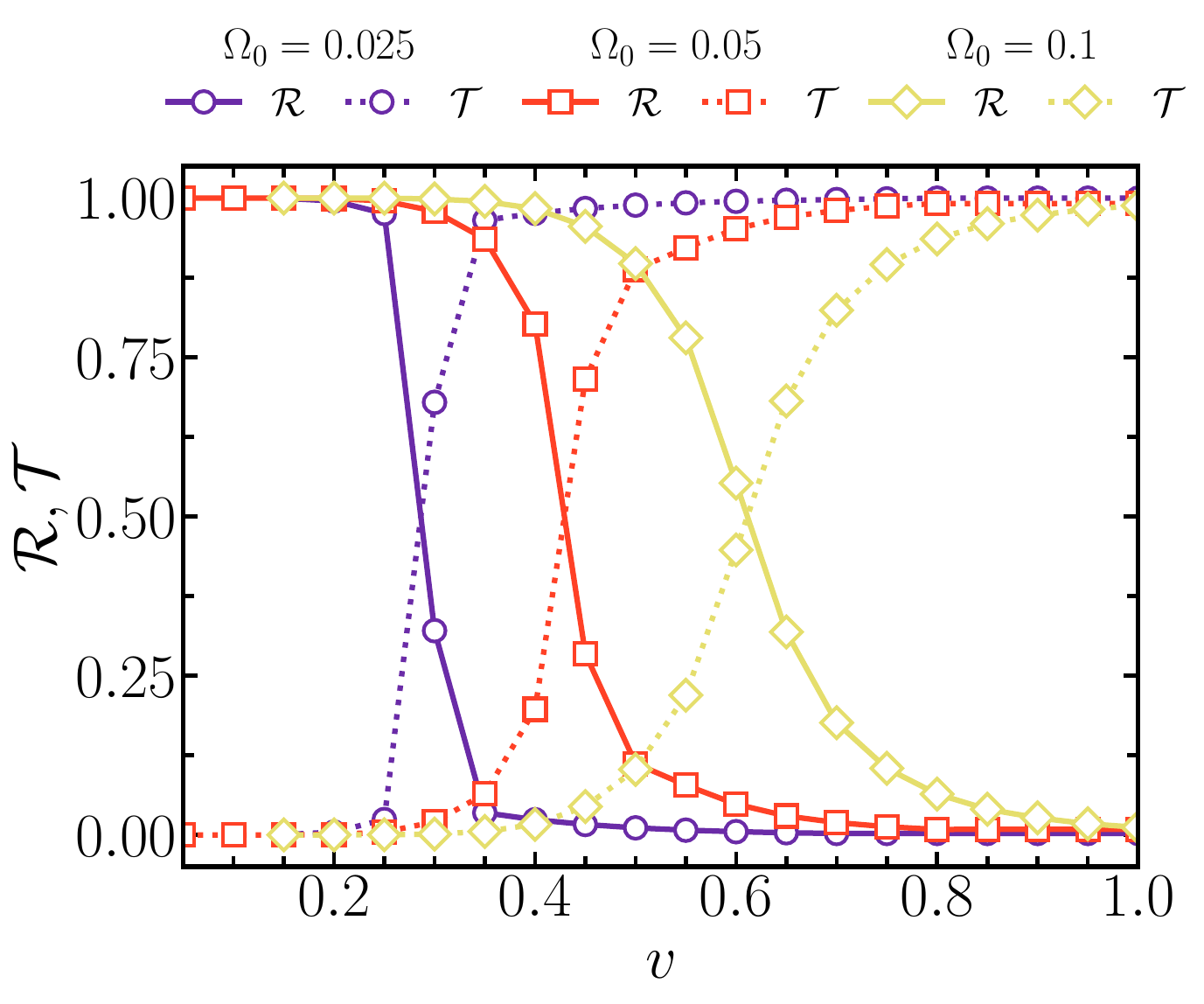}
    \caption{The reflection and transmission coefficients $\mathcal{R}$ and $\mathcal{T}$ as a function of velocity $v$ for $\Omega_0 = 0.025 ~(\rm purple), 0.05 ~(red), $ and $0.1$ (yellow). With increase in velocity, $\mathcal{R}$ decreases monotonically and $\mathcal{T}=1-\mathcal{R}$ increases accordingly, with $\mathcal{R}=\mathcal{T}=0.5$ at  $v_c \approx 0.26$~(purple), $0.42$~(red), and $0.61$~(yellow). Other parameters are kept at $g_{\upar \upar} = g_{\dar \dar} = g_{\upar \dar} = g_{\dar \upar} = -1.0$, and $\Omega_{0} =0.05.$}
    \label{fig:RT-mnkv}
\end{figure}

Next, we obtain the values for soliton reflection and transmission coefficients calculated at a particular time,  $t=2\tau$, as $\mathcal{R} = N_{L}(2\tau)$ and  $\mathcal{T} = N_{R}(2\tau)$, corresponding to the population at the left and the right subspaces, respectively, and calculate their dependence on the soliton velocity. Fig.~\ref{fig:RT-mnkv} represents the threshold value of velocity at which the reflection $\mathcal{R}$ and transmission $\mathcal{T}$ coefficients become comparable with each other ($\mathcal{R} \approx \mathcal{T}$). Further increase in $\Omega_{0},$ leads to increase the $v_c$. Interestingly, the dependence of $v_{c}$ on $\Omega_0$ is analogous to the conventional quantum mechanics formula $v_{c}=2\sqrt{\Omega_{0}}$, beyond which the transmission occurs. 

\begin{figure}[!htp]
    \centering
    \includegraphics[width=\linewidth]{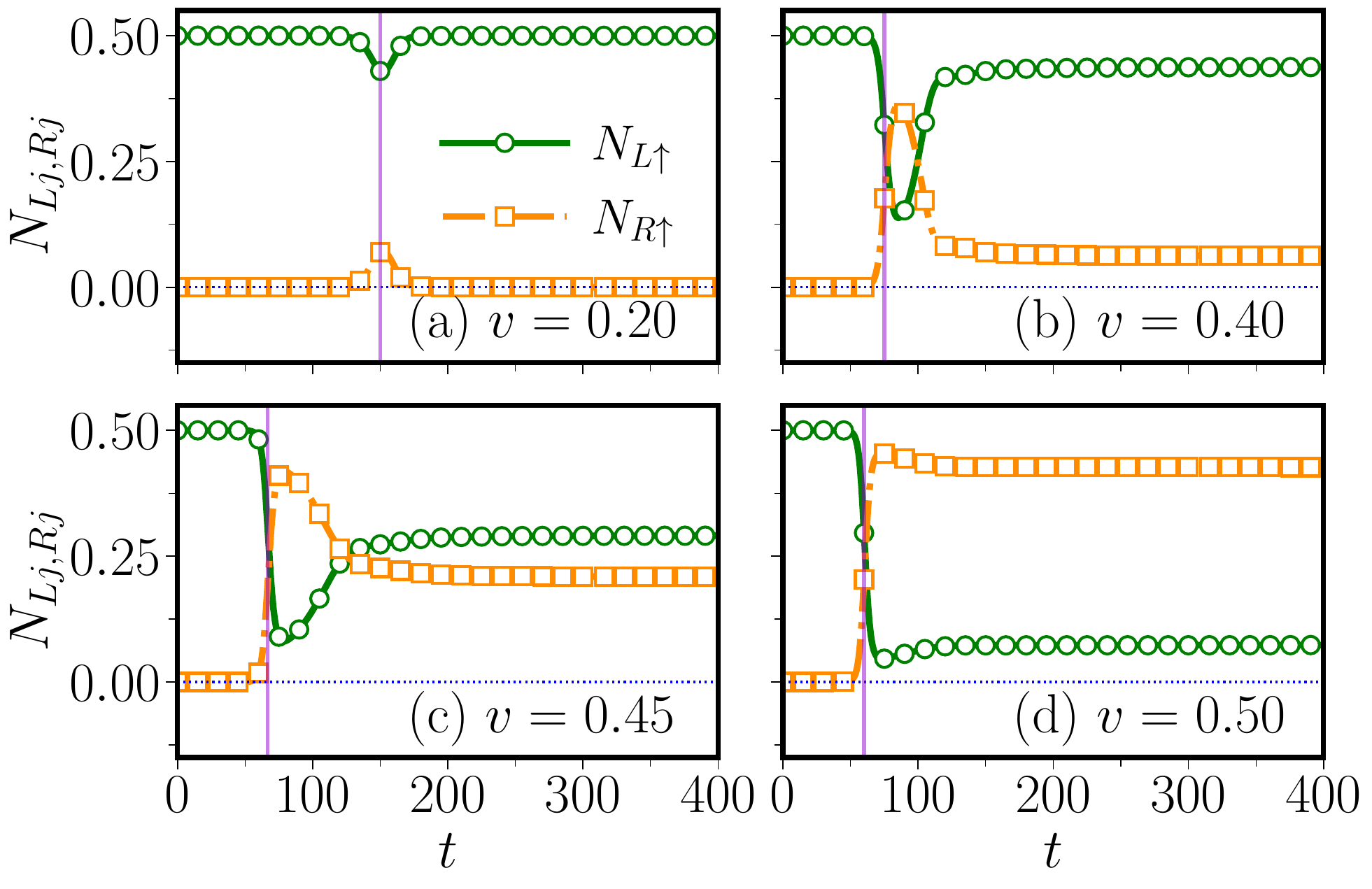}
    \caption{Time-dependent populations  $N_{L\upar}(t)$, and $N_{R\upar}(t)$ for different velocities: (a) $v= 0.2$, (b) $v = 0.4$, (c) $v = 0.45$, and (d) $v = 0.5$. The vertical solid line (violet) shows the time $\tau = |x_0|/v$ at which the soliton interacts with the barrier. For $v = 0.2$, the $N_{L\upar}$ remains constant approximately at $0.5$ and $N_{R\upar}$ is fixed at $0$, depicting zero transmission probability. For other cases, the value of $N_{L\upar}(2\tau) < 0.5$, and $N_{R\upar}(2\tau) > 0$ represent the partial reflection and transmission of the soliton. Nonmonotonic time dependence of $N_{L\upar}$ and $N_{R\upar}$ in the vicinity of $\tau$ partially corresponds to the quantum backflow effect \cite{Palmero:2013,Mardonov:2014,Barbier:2023}. Other parameters are same as Fig.~\ref{fig:den-mnkv}.}
    \label{fig:ref-tran-mnkv}
\end{figure}

For a quantitative analysis, in Fig.~\ref{fig:ref-tran-mnkv} we plot the $N_{Lj}$ and $N_{Rj}$ as a function of time for the same velocities shown in Fig.~\ref{fig:den-mnkv}. For $v = 0.2$, the coefficients remain almost constant at $N_{Lj} \approx 0.5$, and $N_{Rj} \ll 1$ with time, while a small variation at $t  = \tau$ demonstrates an instantaneous change in the soliton profile immediately after the collision at $x=0$. Conversely, for $v = 0.4$, the $N_{Lj}$ shows a sharp cross-over with $N_{Rj}$ defining the exchange of atoms between components following the collision. Further increase at $v = 0.45$, results in a gradual fall in $N_{Lj}$ and increase in $N_{Rj}.$ However, $N_{Lj}$ overcomes the $N_{Rj}$ at longer time, depicting the fact that the $L$ subspace has a slightly larger population compared to the $R$ subspace during the transmission. For a larger velocity, the difference between $N_{Rj} - N_{Lj} \sim N_{Rj}$ is dominantly defined by the population in $R$ subspace, where the other subspace occupation is small revealing the increase of transmission probability.

\begin{figure}[!htb]
    \centering
    \includegraphics[width=\linewidth]{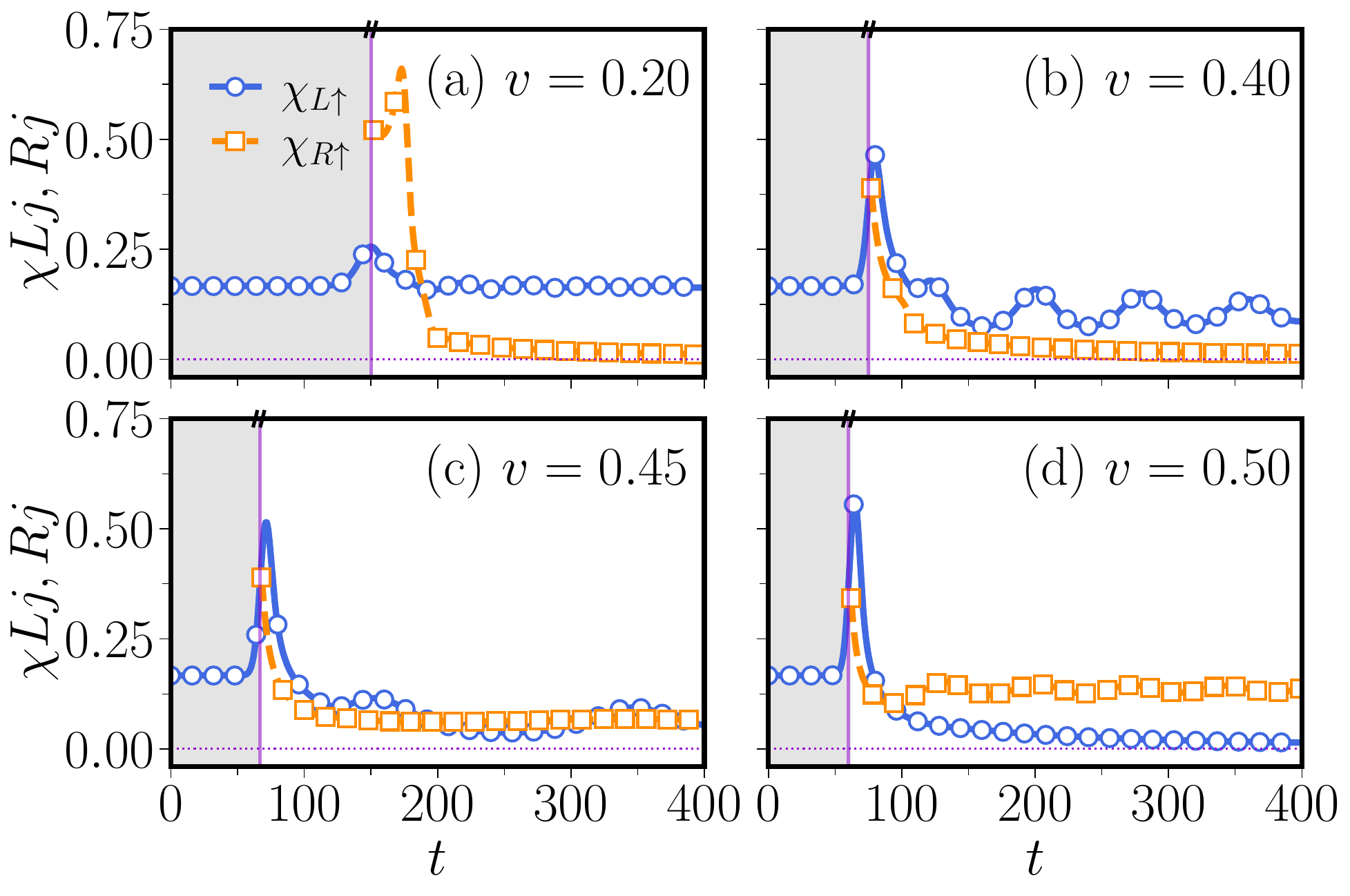}
    \caption{Temporal evolution of the IPRs, $\chi_{Lj}(t)$ and $\chi_{Rj}(t)$, for incident velocities (a) $v=0.2$, (b) $v=0.4$, (c) $v=0.45$, and (d) $v=0.5$. The remaining parameters are the same as those in Fig.~\ref{fig:den-mnkv}. The gray shaded region ($t<\tau$) denotes the interval where $\chi_{Rj}$ is undefined due to the negligible population in the $R$ subspace ($N_{Rj}\ll 1$).}  
    \label{fig:IPR-ref-tran-mnkv}
\end{figure}

Figure~\ref{fig:IPR-ref-tran-mnkv} represents the IPR as a function of time in different subspaces as, $\chi_{Lj}$ and $\chi_{Rj}$, for the same set of incident velocities. The $\chi_R$ is not very well-defined for $t \leq \tau$ (indicated by the shaded gray region) due to the negligible population in that subspace before collision, and therefore, the quantity is not shown in that region. The finite values of $\chi_{Lj}$ and $\chi_{Rj}$ for $t > \tau$ indicate finite atomic populations in the left ($L$) and right ($R$) subspaces, respectively. For $v = 0.40$ (in Fig.~\ref{fig:IPR-ref-tran-mnkv}(b)), $\chi_{L\upar}$ exhibits small-amplitude oscillations, reflecting the partial transmission of the soliton with time in the $R$ subspace. Upon increasing the incident velocity to $v=0.45$, the IPRs of the two subspaces become nearly comparable for $t > \tau$. In fact, for higher velocities, $\chi_{Rj}$ exceeds $\chi_{Lj}$ for $t \gg \tau$, demonstrating that the transmitted component becomes more localized than the reflected one as transmission dominates over reflection. 

An interesting feature of the soliton transmission indicated by behavior of the IPR is its jam in the transmission process seen as a considerable increase in the IPR, $\chi_{\rm ex}$ at the exit from the Rabi inhomogeneities. This jam appears since different points of the soliton have different velocities with those at larger $x$ have smaller velocities since they are closer to larger values of $\Omega(x).$ For this reason, the transmitted soliton has a higher IPR than the reflected one with $\chi_{\rm ex}>\chi_{Lj}(0).$

Oscillation in the condensate density, very clearly seen in Figs. \ref{fig:den-mnkv}(b) and  \ref{fig:IPR-ref-tran-mnkv}(b), present another characteristic feature of the reflection and transmission.  As these Figures show, the population of the condensate after reflection is ${\mathcal R}\approx 0.75$, while the IPR after the exit, where the decrease is transferred into oscillating behavior, $\chi_{\rm tr}$, remains almost the same as the initial $\chi_{Lj}(0).$ Thus, the attractive nonlinearities at the decreased density become insufficient to keep the value of $\chi_{\rm tr}\approx\chi_{Lj}(0),$ the width tends to increase and the oscillations start to dynamically compensate for this imbalance.  A simple variational approach analysis shows that the oscillation amplitude is approximately $\chi_{\rm tr}\left(1-{\mathcal R}\right)$ and the period $8\sqrt{3}\pi/g^{2}{\mathcal R}^{2},$ in good agreement with the numerical results. A similar pattern of oscillations is seen in \ref{fig:IPR-ref-tran-mnkv}(c), where decreasing in ${\mathcal R}$ and  $\chi_{\rm ex}$ produce oscillations with a smaller amplitude and a larger period.

\begin{figure}[!htp]
    \centering
    \includegraphics[width=1.05\linewidth]{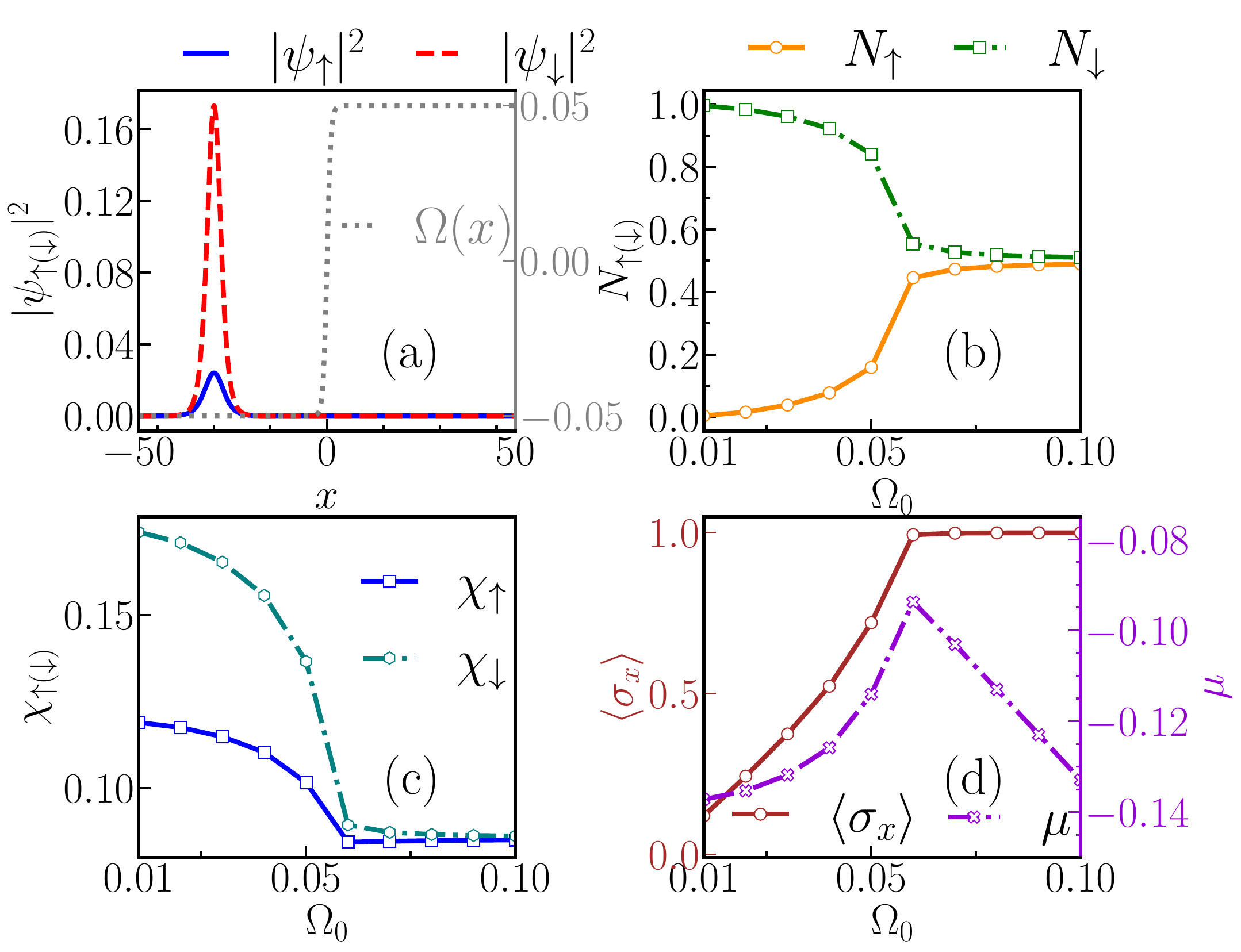}
    \caption{(a) Ground state density profiles for $\Omega_{0} = 0.05$ for $x_0 = -30,$  where the Rabi field is $-\Omega_{0}.$ On the right side axis, the spatial dependence of $\Omega(x)$ is shown.   Variation of (b) population $N_{\upar (\dar)},$ (c) IPR $\chi_{\upar(\dar)}$, and (d) $\langle \sigma_x \rangle$ and chemical potential $\mu$ as a function of $\Omega_0.$ The increase in $\Omega_0$ equalizes the population and shape of the condensate by transferring atoms from $\dar$ to $\upar$ component, as demonstrated by tending $\langle \sigma_x \rangle$ to one.  The interaction parameters are $g_{\upar \upar} = -1.0, g_{\dar \dar} = -1.05$, and $g_{\upar\dar} = g_{\dar\upar} = 0.$}
    \label{fig:density}
\end{figure}

\subsection{Soliton with lifted Manakov symmetry}
\begin{figure*}[!htp]
\centering
 \includegraphics[width=\linewidth]{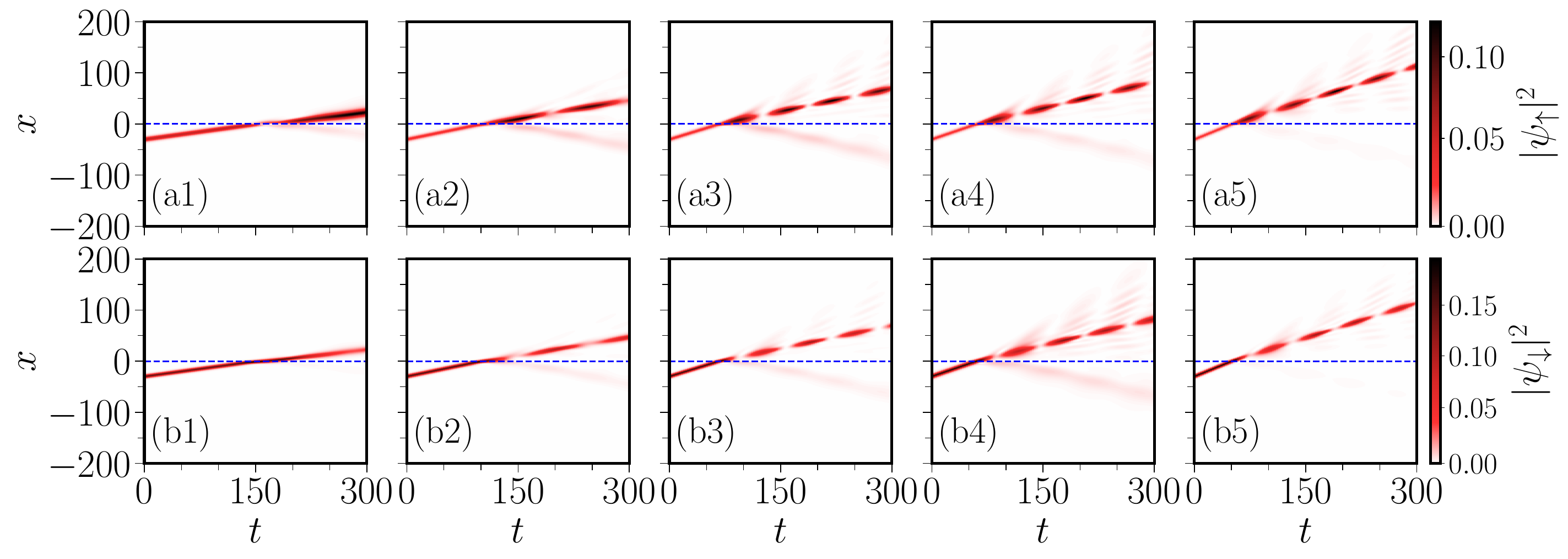}
\caption{Condensate densities $|\psi_\upar|^2$ (top), and $|\psi_\dar|^2$ (bottom) for different velocities:(a1),(b1) $v = 0.2$, (a2),(b2) $v = 0.3$, (a3),(b3) $v = 0.45$, (a4),(b4) $v = 0.5$, and (a5),(b5) $v = 0.6$. For $v = 0.2$ the soliton is uniformly transmitted through the $x = 0$ barrier. As the velocity increases, the transmission probability increases while the reflection probability gradually decreases. The imbalance in population of spin components occurring at transmission from $-\Omega_0$ to $\Omega_0$ produces population probabilities. Other parameters are $g_{\upar \upar} = -1.0, g_{\dar \dar} = -1.05, g_{\upar \dar} = g_{\dar \upar} = 0, \Omega_0 =0.05$, and initial $x_{0} = -30$.}
\label{fig:t-x-density-non-int}
\end{figure*}

Although the Manakov symmetry has a single realization with all the nonlinear coupling constants being equal, it can be lifted in many different ways, still preserving certain symmetries of the spin-1/2 condensate. Here, we consider two realizations of the lifted Manakov symmetry.

We begin with the $g_{\upar \upar} = g_{\dar \dar} = 0$, $g_{\upar \dar} = g_{\dar \upar}$ realization. In this case the initial ground state  $\psi_{{\rm gs},\upar}(y)=\psi_{{\rm gs},\dar}(y)={\sqrt{|g_{\dar \upar}|}}/{4\rm cosh}(yg_{\dar \upar}/4)$ corresponds to $\langle \sigma_{x}\rangle=1,$ the same as in the Manakov symmetry. This similarity results in a similar scattering process, and therefore, we will not consider it in detail.

A more interesting realization of lifted Manakov symmetry, clearly demonstrating the effect of nonlinearities, is $g_{\upar \upar} = g_{\dar \dar},$ $g_{\upar \dar} = g_{\dar \upar} = 0.$  Here, the ground state is degenerate, corresponding to zero occupation either of spin-up or spin-down state. To lift this degeneracy we introduce a weak disbalance $g_{\dar \dar} = g_{\upar \upar}(1+\epsilon)$ with $0<\epsilon\ll\,1.$

To begin with, we consider the ground state of a soliton located at $x_{0}<0$ with $|x_{0}|\gg\, 1$ in the Rabi field $-\Omega_{0}.$ In this realization in the $\Omega_{0} = 0$ limit, the spin-up state is empty while the spin-down state is fully populated. In the $\epsilon=0^{+}$ limit the corresponding ground state energy becomes $-g^{2}/24$ and the chemical potential $\mu_{s}=-g^{2}/8.$ Thus, already $\Omega_{0}\sim |\mu_{s}|\ll g^{2}$ can have a strong impact on the soliton.

The trial function for studying the ground state can be written as 
\begin{equation}
%{\bm\psi}(y)= \left[\frac{\sqrt{N_{\upar}}}{\sqrt{2w_{\upar}}}\frac{1}{\cosh(y/w_{\upar})},\frac{\sqrt{N_{\dar}}}{\sqrt{2w_{\dar}}}\frac{1}{\cosh(y/w_{\dar})} \right]^{\rm T},
\psi_{{\rm gs},j}(y)= \frac{\sqrt{N_{j}}}{\sqrt{2w_{j}}}\frac{1}{\cosh(y/w_{j})}.
\end{equation}
At $\Omega_{0}\ll |\mu_{s}|$ we obtain $N_{\upar}=36\Omega_{0}^{2}/g^{4},$  $w_{\dar}=2/|g|$ ($\chi_{\dar}=|g|/6$), $w_{\upar}=w_{\dar}(1+\zeta),$  where $g-$independent $\zeta$ denotes the relative broadening of the spin-up component. This $g-$ independence is due to the fact that at $N_{\upar}\ll\,1$ nonlinearities do not play a significant role in the ${\upar}$ component since the term $g_{\upar \upar} \vert \psi_{\upar} \vert ^{2}$ in Eq. \eqref{eqn1(a)} is negligibly (by parameter $\sim\Omega_{0}^{2}/g^{4}$) small  compared to the kinetic and Rabi energies. Thus, $\chi_{\upar}$ is determined by minimization of the sum of competing kinetic (decreasing with the increase of $w_{\upar}$) and Rabi (increasing with the increase of $w_{\upar}>w_{\dar}$ due to a smaller overlap of the $\upar$ and $\dar$ components) energies.  Notice that despite a large difference in $N_{\dar}$ and $N_{\upar},$ the corresponding $w_{\dar}$ and $w_{\upar}$ are close due to the Rabi-induced mutual localization of the spin components \cite{Santos:2021,Sarkar:2025b}. The corresponding Rabi energy in Eq. \eqref{eq:ER} becomes $E_{\Omega}=-12f(\zeta)\Omega_{0}^{2}/g^{2},$ where 
\begin{equation}
f(\zeta) = \frac{1}{2\sqrt{w_{\upar}w_{\dar}}}
\int_{-\infty}^{\infty}\frac{dy}{\cosh(y/w_{\upar})\cosh(y/w_{\dar})} \leq 1, 
\end{equation}
corresponds to the overlap of the non-spatially identical spin components. 

This response to the Rabi field $\Omega\sigma_{x}$ shows that this setting of nonlinearities can be mapped onto a Rabi-like field along the $z-$axis $B\sigma_{z}$ with $B=g^{2}/12.$ Thus, the effective field acting on the soliton spin has the form $\left(-\Omega_{0},0,B\right)$, producing a system similar to those demonstrating the Van Vleck paramagnetism \cite{Fazekas:1999}. In the opposite case $\Omega_{0}\gg |\mu_{s}|$ we have $w_{\upar}=w_{\dar}=4/|g|$ and $N_{\dar}-N_{\upar}\approx \epsilon g^{2}/24\Omega_{0}.$ 

In Fig.~\ref{fig:density}(a) we show the ground state soliton profile at $x_{0} = -30$ and $\Omega_{0} = 0.05.$ Figure ~\ref{fig:density}(b) represents that  $N_{\dar}>N_{\upar}$ ($N_{\upar}=0$ at $\Omega_{0}=0$) with $N_\upar \approx N_\dar$  for $\Omega_0>0.05.$ Correspondingly, the increase in $\Omega_0$  results in lowering the IPR to $\chi_{\dar}=\chi_{\upar}=1/12$ (Fig.~\ref{fig:density}(c)). Notice that Fig.~\ref{fig:density}(c) shows that in the limit of $\Omega_{0}=0,$ one obtains $\chi_{\dar}(\Omega_{0})\approx 1.46\chi_{\upar}(\Omega_{0}),$ 
 consistent with the $\zeta\approx 0.32$ obtained by the variational approach, which somewhat underestimates the width of the spin-$\upar$ component since the Zeeman-induced localization effect is weaker than that induced by the nonlinearities. The chemical potential $\mu$ decreases in strong fields to $d\mu/d\Omega_{0}=-1$ (Fig.~\ref{fig:density}(d)) with $\langle \sigma_x \rangle \approx 1$ at $N_\upar \approx N_\dar.$

Next, in Fig.~\ref{fig:t-x-density-non-int}, we show the condensate densities $|\psi_{\upar}|^2$ (upper panel), and $|\psi_{\dar}|^2$ (lower panel) in the $(t,x)$ plane for different velocities $v.$ Here the soliton gets transmitted uniformly to $R$ subspace even with a smaller velocity compared to Manakov symmetry, as shown in figures (a1,a2) and (b1,b2) for $v = 0.2.$ As the velocity increases [(a3)–(a5) and (b3)–(b5)], periodic modulation develops with the emission of weak dispersive wave packets after transmission. This periodic modulation is much stronger and has a more complex pattern than in the case of Manakov symmetry. This difference occurs since the population exchange between the components is caused by the Rabi field, in addition to the density oscillations due to the imbalance between the state occupation and the widths, as in the Manakov symmetry. The radiation becomes stronger at larger velocities, leading to a sequence of localized low-density maxima along the soliton path. Despite the enhanced radiation, the soliton remains stable and continues its propagation while the evolution of the spin components exhibits qualitatively similar behavior.

 For lifted Manakov symmetry, the mismatch between the instantaneous soliton configuration and the adiabaticity-produced state results in exciting its internal modes, seen as spin-dependent nonlinear oscillations in the condensate density and density radiation. With increasing incident velocity, the soliton spends less time adapting to the local change in $\Omega(x),$ leading to more pronounced density modulation due to stronger non-adiabatic effects. The sequence of density maxima and minima along the trajectory reflects the repeated exchange of energy and population between the spin-density modes.

	\begin{figure}
			\centering
			\includegraphics[width=\linewidth]{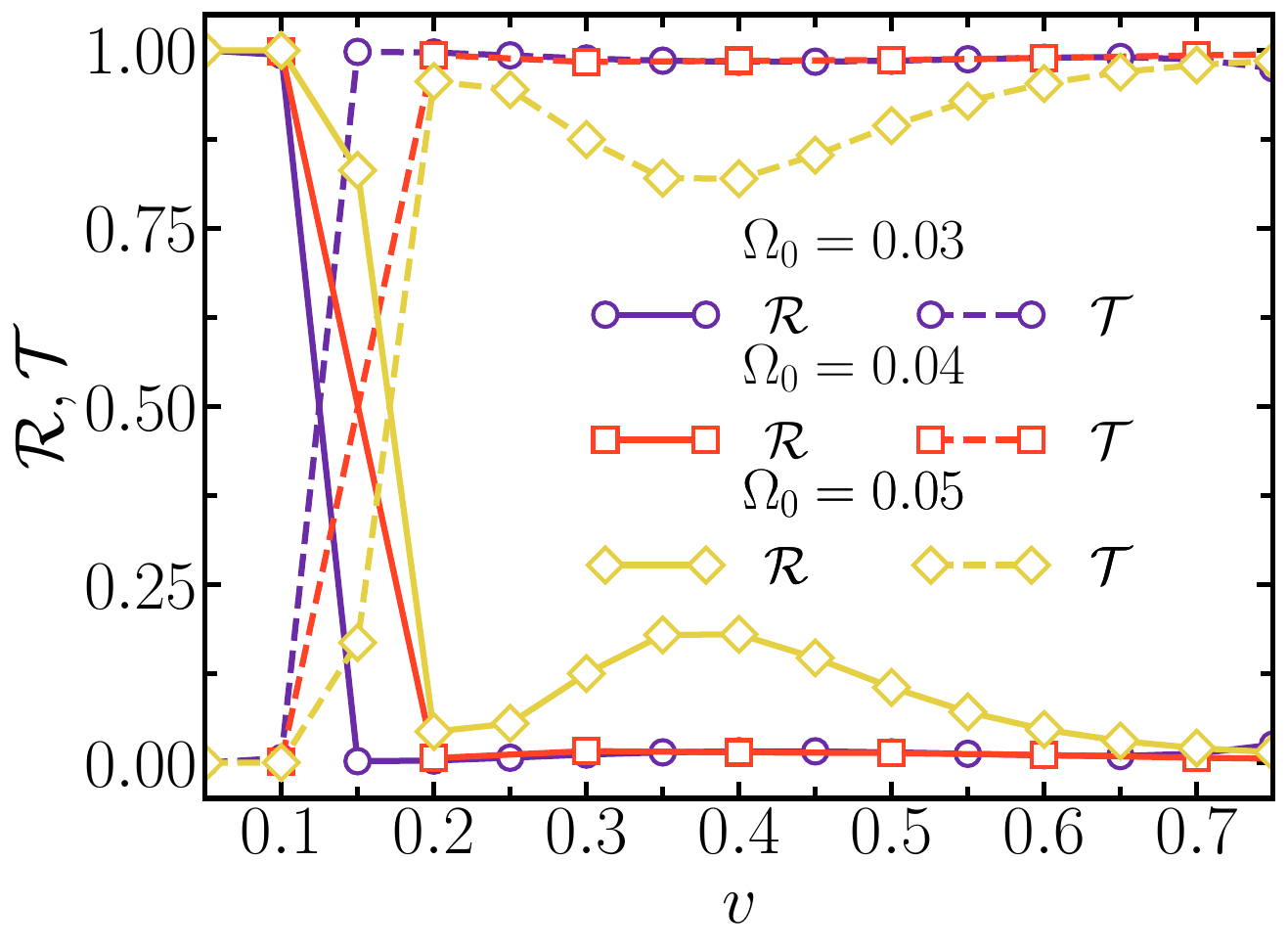}
			\caption{The reflection and transmission coefficients $\mathcal{R}$ and $\mathcal{T}$ as a function of velocity $v$ for $\Omega_0 = 0.03$ (purple), $0.04$ (red), and $0.05$ (yellow). 
				The critical velocity at which $\mathcal{R} \approx \mathcal{T} \approx 0.5$ are $v_c \approx 0.12,$ $0.15$, and $0.17$, respectively. Here are $g_{\upar \upar} = -1.0, g_{\dar \dar} = -1.05, g_{\upar \dar} = g_{\dar \upar} = 0$. }
			\label{fig:RT-bmnkv}
		\end{figure}
		
Figure~\ref{fig:RT-bmnkv} illustrates the dependence of the reflection and transmission coefficients on the initial velocity $v$ for different Rabi couplings. Similarly to the Manakov symmetry, the critical velocity increases with $\Omega_{0},$ due to enhanced effective barrier height, which requires a larger soliton kinetic energy to overcome the barrier. However, the critical velocity for the lifted Manakov symmetry is lower than that in the Manakov case, as evident from the comparison with Fig.~\ref{fig:RT-mnkv}. For $\Omega_0 = 0.05$, the amplitudes of the coefficients exhibit oscillatory behavior, which originates from periodic oscillations of the population after transmission to the $R$ subspace. For other values of $\Omega_0$, the coefficients remain approximately constant after transmission.

The critical velocity is strongly influenced by the symmetry of spin-dependent interactions. Since in the Manakov case the soliton $\langle\sigma_{x}\rangle=1$ remains constant, a velocity $v>2\sqrt{\Omega_{0}}$ is required for efficient transmission. In contrast, for the lifted Manakov symmetry, the effective barrier depends on the details of the soliton spin state. In this case, when the soliton approaches the barrier, the spin-dependent interactions modify the relative phase and population imbalance between components. This modification enables the soliton to adapt more efficiently to the variation of Rabi coupling, thereby reducing the effective height of the barrier. Therefore, the lifted Manakov symmetry provides additional internal channels through which the soliton kinetic energy leads to spin and density redistribution, permitting the soliton to cross the barrier efficiently at a lower critical velocity for transmission.    

\begin{figure}
    \centering
    \includegraphics[width=\linewidth]{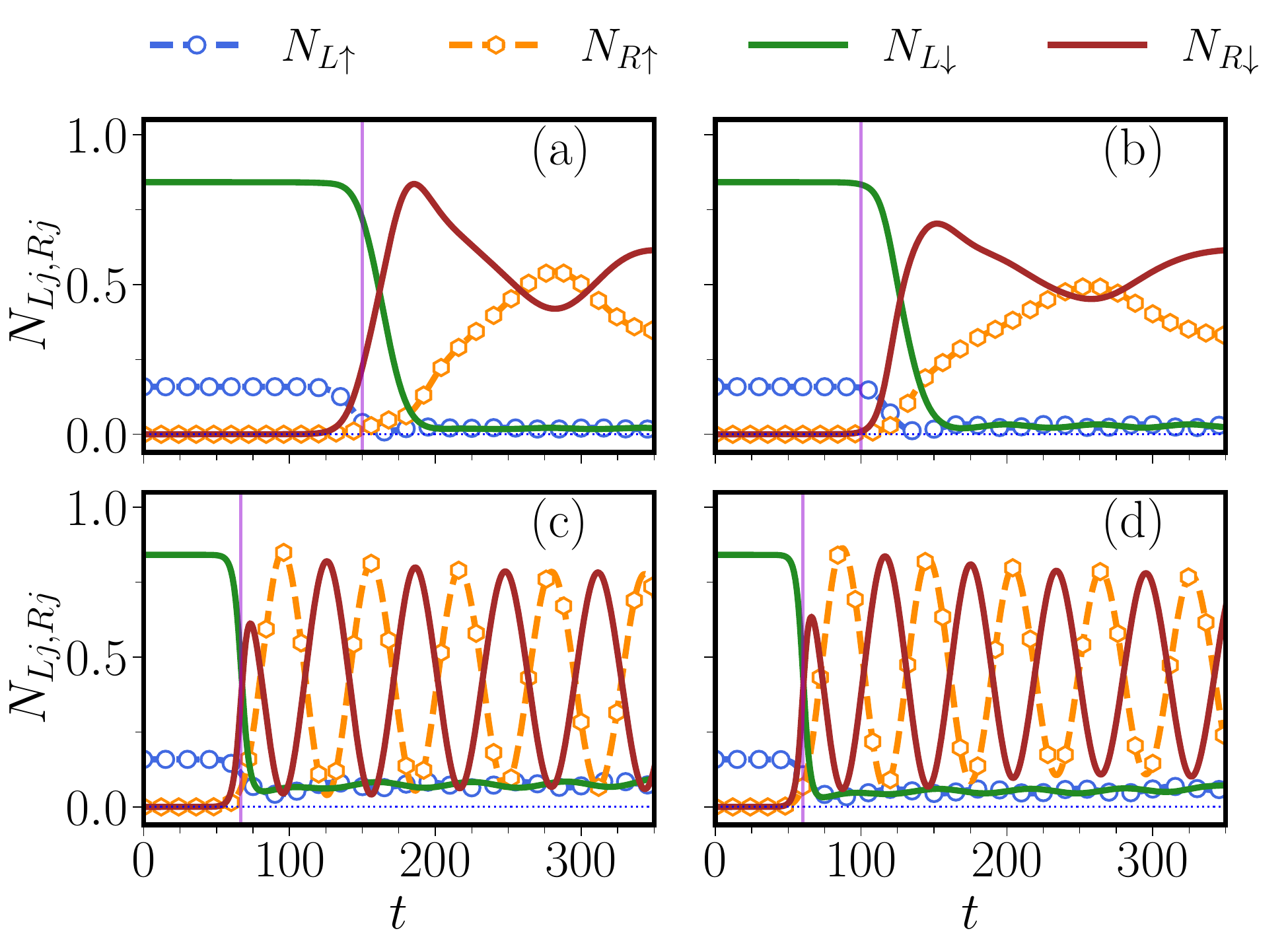}
    \caption{Time evolution of the spin-projected populations in $L$, and $R$ subspaces, $N_{L {\uparrow(\downarrow)}}(t)$ and $N_{R {\uparrow(\downarrow)}}(t)$, for different incident velocities: (a) $v = 0.2$, (b) $v = 0.3$, (c) $v = 0.45$, (d) $v = 0.5.$ In all cases $N_{Lj}$ remain finite for $t \leq \tau$, while $N_{Rj}$ stay nearly zero. For $t > \tau$, $N_{Lj}$ rapidly vanish and $N_{Rj}$ develop periodic oscillations in time. } 
    \label{fig:ref-trans-b-mnkv}
\end{figure}

To gain further insight into the collision dynamics, Fig.~\ref{fig:ref-trans-b-mnkv} presents the time evolution of the spin populations, $N_{Lj}$ and $N_{Rj}$, in the left ($L$) and right ($R$) subspaces, respectively. For relatively low incident velocities ($v=0.2$ and $0.3$), the soliton is predominantly reflected after interacting with the local Rabi-coupling region. Consequently, for $t>\tau$, the populations in both spin components remain nearly constant, indicating negligible population exchange between the spin-$\upar$ and spin-$\dar$ states [Figs.~\ref{fig:ref-trans-b-mnkv}(a) and (b)]. At these velocities, the longer interaction time allows the soliton to adjust smoothly to the local Rabi field, resulting in an almost stationary spin composition after the collision.

In contrast, for larger incident velocities with $v\geq 0.45$, the interaction time with the inhomogeneous Rabi coupling region is significantly reduced. As a result, the collision excites coherent internal spin dynamics, leading to a periodic population exchange between the spin-$\upar$ and spin-$\dar$ components and emission of weak dispersive radiation. 

\begin{figure}
    \centering
    \includegraphics[width=\linewidth]{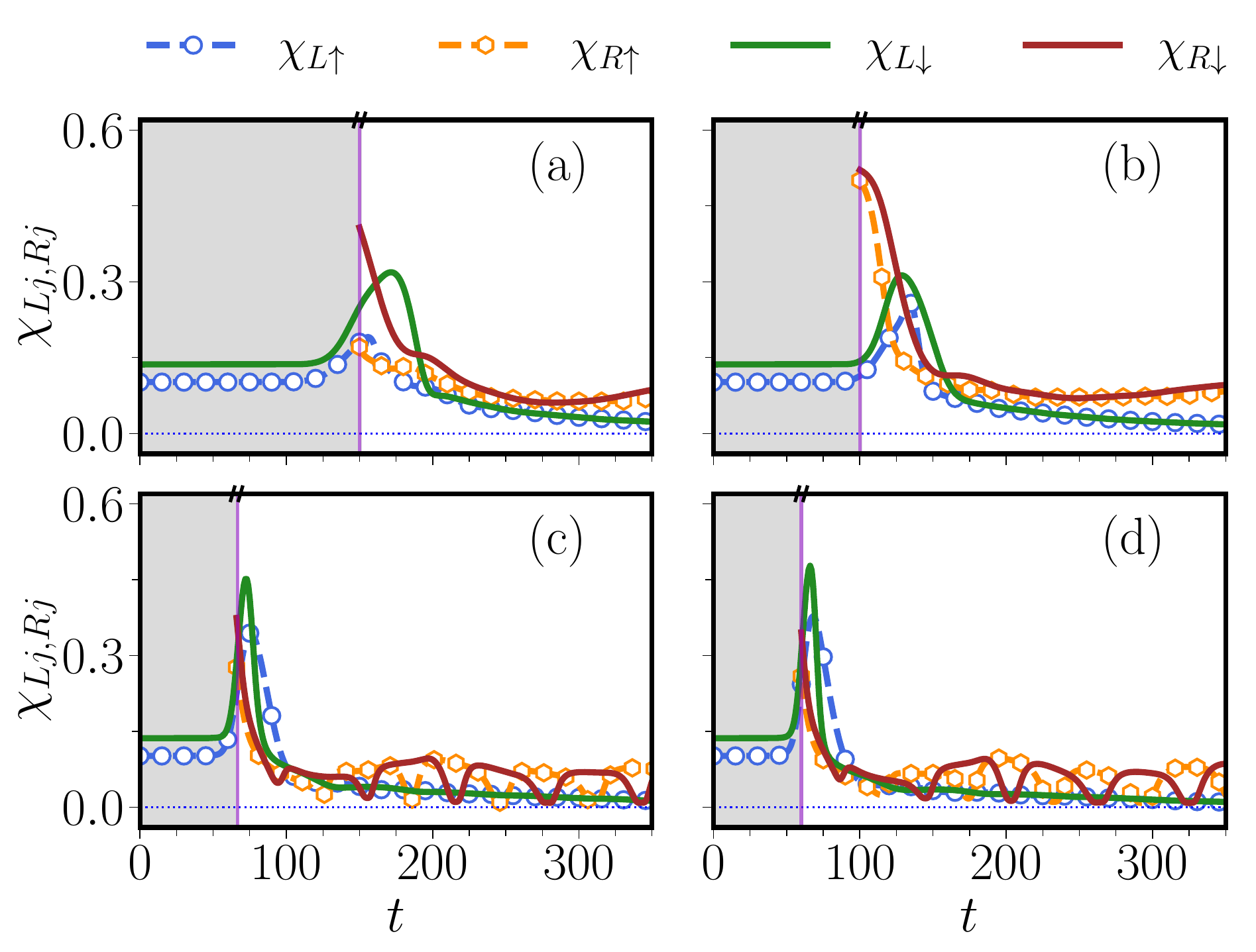}
    \caption{Time evolution of the spin-projected IPR, $\chi_{L{\uparrow(\downarrow)}}(t)$ and $\chi_{R{\uparrow(\downarrow)}}(t)$, for different incident velocities: (a) $v = 0.2$, (b) $v = 0.3$, (c) $v = 0.45$, (d) $v = 0.5.$ For $v = 0.2, 0.3$, the  $\chi_{R\upar(\dar)}$ nearly saturates, while for higher velocities $v\geq 0.45$ IPR's in the $R$ subspace exhibit oscillation over time.} 
    \label{fig:IPR-b-mnkv}
\end{figure}

Figure~\ref{fig:IPR-b-mnkv} depicts the time evolution of $\chi_{Lj}(t)$ and $\chi_{Rj}(t).$ As the soliton reaches the Rabi barrier at $t=\tau$, all spin-projected IPRs exhibit pronounced peaks due to the transient compression of the soliton. For lower incident velocities [Figs.~\ref{fig:IPR-b-mnkv}(a) and (b)], the post-collision IPRs decay smoothly and attain nearly constant values, indicating that the reflected and transmitted wave packets preserve their localized solitonic nature with minimal internal spin dynamics. In contrast, for higher velocities [Figs.~\ref{fig:IPR-b-mnkv}(c) and (d)], the IPRs display damped oscillations after the collision, reflecting coherent Rabi-induced population exchange between the spin components accompanied by weak radiation emitted during transmission. The gradual reduction of the oscillation amplitude at longer times signifies the relaxation of the emitted radiation, while the finite asymptotic values of the IPRs confirm that the transmitted soliton remains spatially localized.


\subsection{Large soliton velocity: quenched Rabi coupling}
\begin{figure}[!htp]
    \centering
    \includegraphics[width=\linewidth]{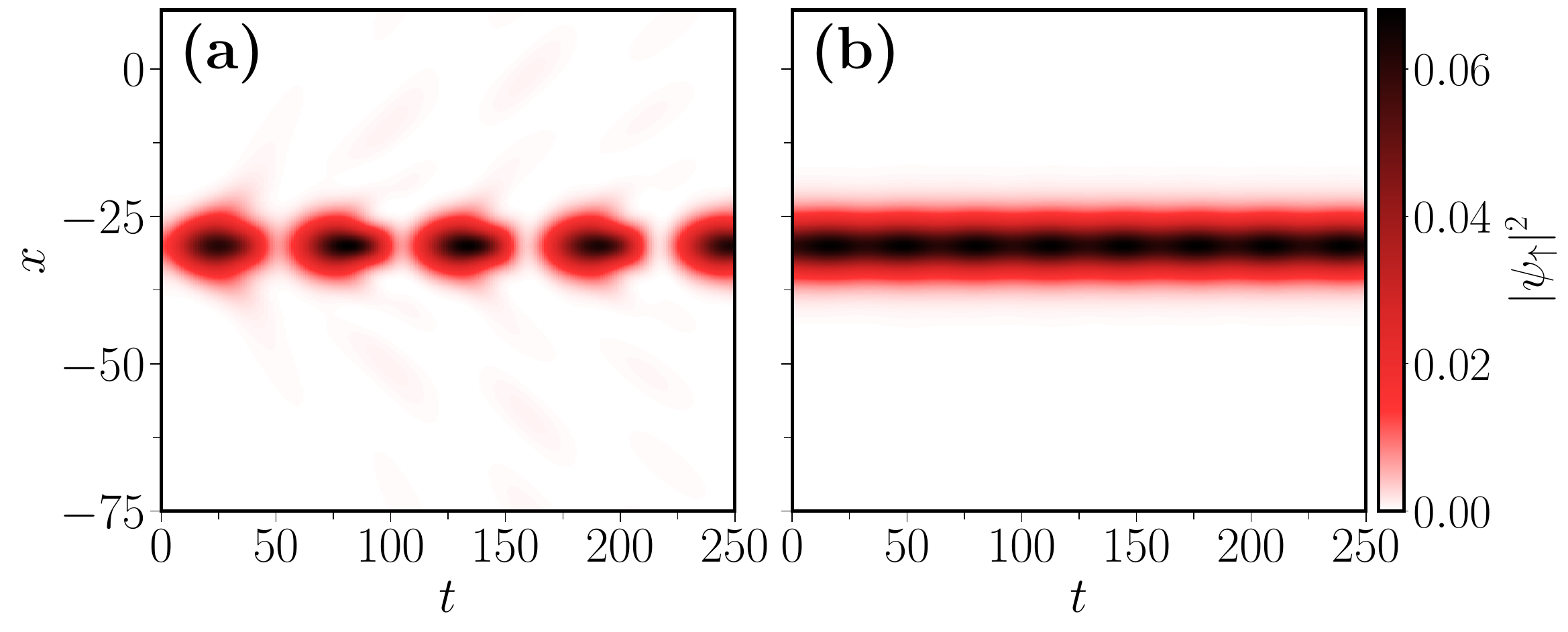}
    \caption{Spatiotemporal evolution of the $|\psi_{\uparrow}|^2$ in $(t,x)$ plane, followed by a sudden quench of $\Omega_0$ as (a) $-0.05 \rightarrow 0.05$, and (b) $-0.08 \rightarrow 0.08$. Upon quenching $\Omega_0$ at $0.05$ (a), the population exchange occurs between the spin components. The interaction parameters are $g_{\upar \upar} = -1.0, g_{\dar \dar} = 1.05g_{\upar \upar}, g_{\upar \dar} = g_{\dar \upar} = 0$.}
    \label{fig:t-x-density-quench}
\end{figure}
In the previous subsection, we investigated the soliton dynamics induced by motion across the spatial Rabi barrier. Upon crossing the barrier the soliton encounters a change of Rabi coupling between $-\Omega_0$ to $\Omega_0$ from $x< 0$ to $x > 0.$  During the transition, the competition between soliton's kinetic energy and the Rabi barrier governs its reflection and transmission, as described above. 

Under certain conditions soliton can propagate intact through the Rabi barrier, with the change in sign of the Rabi coupling being analogous to its quenching. This quench-based approach can be a powerful tool of probing coherent soliton spin dynamics after motion through the barrier. 

For a small $\Omega_{0}\ll |\mu_{s}|$ a quench-like dynamics produced regime can be realized for sufficiently large velocities $v.$ This regime can occur if the passing time through the barrier $W/v\ll T_{R}/4,$ where $W$ is the width of the Rabi barrier (here we can take $W=2$) and $T_{R}=12\pi/g^{2}$ is the characteristic spin precession period in the field $B=g^{2}/12.$  In this realization the field $\left(-\Omega_{0},0,B\right)$ rapidly changes to $\left(\Omega_{0},0,B\right)$, and the soliton spin starts to precess around the vector ${\mathbf n}=\left(\Omega_{0},0,B\right)/\sqrt{\Omega_{0}^{2}+B^{2}}$ with the rate close to $2B$ and  changes in $N_{\upar}$ scaling as $\max(N_{\upar})\approx 576\Omega_{0}^{2}/g^{4}.$ It corresponds to the effective barrier $2|E_{\Omega}|$ with the critical $v_{c}=2\sqrt{|E_{\Omega}|}=4\sqrt{3}\Omega_{0}/|g|,$ where we took $f(\zeta)=1$ for simplicity. Therefore, to observe the spin quenching regime, the velocity should simultaneously satisfy two conditions: (1) $v\gg 4\sqrt{3}\Omega_{0}/|g|$ and (2) $v\gg Wg^{2}/3\pi,$ as can be presented by defining a critical velocity as:
\begin{equation}
   v\gg\,v_{c}=\max\left(4\sqrt{3}\frac{\Omega_{0}}{|g|}, \frac{Wg^{2}}{3\pi}\right). 
   \label{eqn:v_crit}
\end{equation} 
Just the condition (2) does not ensure realization of the quench regime even at a small $\Omega_{0}$ due to a relatively large prefactor $4\sqrt{3}$ and a small prefactor $1/3\pi$ in conditions (1) and (2), respectively.

For $\Omega_{0}\gg |\mu_{s}|$, the reflection pattern is similar to the Manakov symmetry case since the soliton spin is approximately conserved at $\langle\sigma_{x}\rangle=1$ and the critical velocity becomes $2\sqrt{\Omega_{0}}$ with the quenching realized at $v\gtrsim v_{c}.$

Here we investigate the quenching dynamics of the ground state soliton with the sudden change in $\Omega$ as $-\Omega_0 \rightarrow \Omega_0$. The switching of Rabi coupling is likely analogous to the change in the $\Omega(x)$ slope as the moving soliton crosses $x = 0$ barrier. To capture the dynamics, we calculate the condensate density, populations, and spin polarizations after quenching.   
\begin{figure}[!ht]
    \centering
    \includegraphics[width=\linewidth]{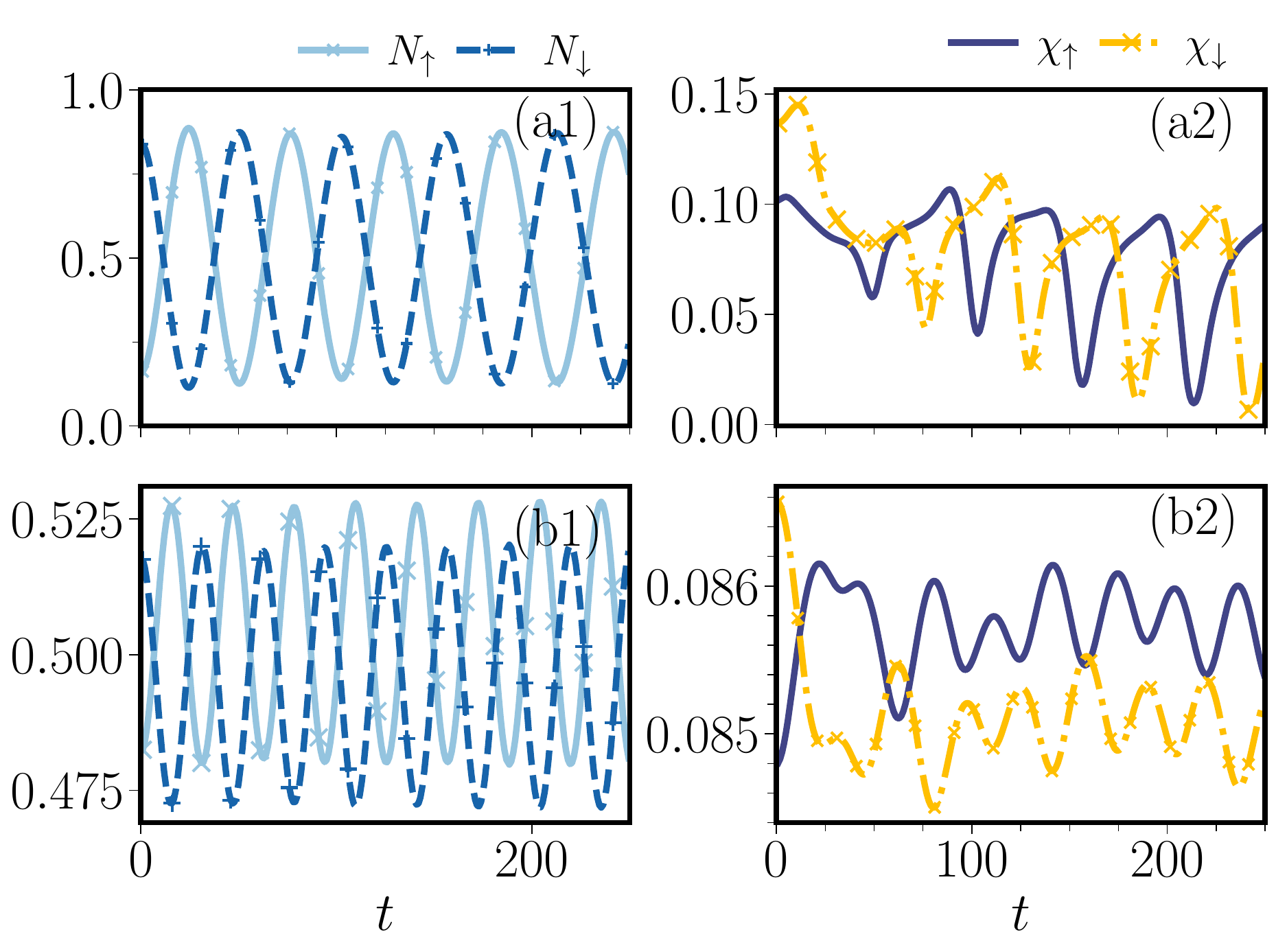}
    \caption{Time evolution of the spin populations, $N_j(t)$ (left panel), and IPR $\chi_j(t)$ (right panel), obtained after quenching the Rabi field as: (a1,a2) $-0.05 \rightarrow 0.05$, and (b1,b2) $-0.08 \rightarrow 0.08$. Coherent population oscillations are prominent for smaller value of $\Omega_{0}.$ Increasing $\Omega_{0}$ progressively suppresses the oscillation amplitude, leading to nearly balanced spin populations. The IPR also decreases over time for $\Omega_{0} = 0.05$, whereas for larger Rabi coupling the IPR remains almost constant. Here, the interaction parameters are the same as in Fig.~\ref{fig:t-x-density-quench}.}
    \label{fig:Nup-Nd-quench}
\end{figure}

\begin{figure}[!ht]
    \centering
    \includegraphics[width=\linewidth]{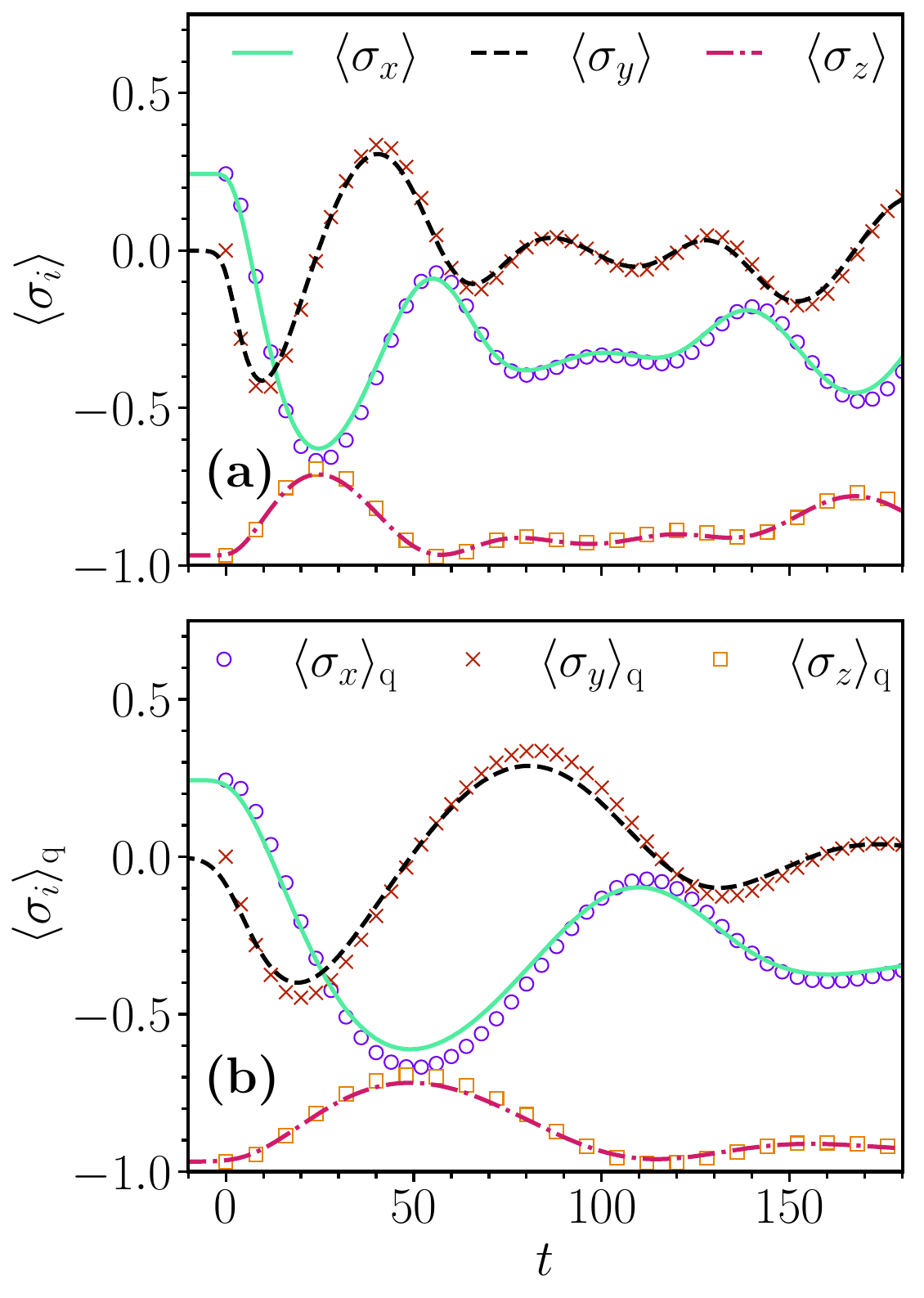}
    \caption{Spin polarization components $\langle \sigma_i \rangle$ obtained from velocity perturbation (lines) compared with the sudden quench of the Rabi coupling from $-\Omega_{0}$ to $\Omega_0$ at $t=0$ (markers). In (a) $g_{\upar \upar} = -1, \Omega_{0} = 0.02, v = 0.84\approx 4 v_{c},$ and (b)  $g_{\upar \upar} = -1/\sqrt{2}, \Omega_{0} = 0.01, v = 0.47\approx 4v_{c}$. Here, $v_{c}$ follows Eq.~\ref{eqn:v_crit}, and other interaction parameters are $g_{\dar \dar} = 1.05 g_{\upar \upar},$ and $g_{\upar \dar} = g_{\dar \upar} = 0$. For velocity-perturbed dynamics, time is measured after $t > \tau$, when the soliton center crosses $x = 0$. In both cases, the $v_c$ is calculated as ${W g^2}/{3\pi}$ which turns out to be  $0.2122$ (in (a)) and $0.1061$ (in (b)). We notice, however, that in these cases  $4\sqrt{3}{\Omega_{0}}/{|g|}$ and ${W g^2}/{3\pi}$ are numerically close, implying that both criteria are of the same importance.   
   }
    \label{fig:spin-pol-Omega-quench}
\end{figure}

Fig.~\ref{fig:t-x-density-quench} shows the evolution of the condensate density $|\psi_\upar|^2$ in the $(t,x)$ plane followed by the quenching of Rabi coupling. For $\Omega_{0} = 0.05$, the density exhibits periodic oscillation over time due to the considerable population imbalance between the components with $N_{\upar}>N_{\dar}.$ However, with the increase to $\Omega_{0} = 0.08,$ the population exchange gets diminished because of equalizing the population between the components. The population exchange is shown in Fig.~\ref{fig:Nup-Nd-quench}, where for $\Omega_0 = 0.05$ (in (a1)), the difference $|N_\upar - N_\dar|$ always remains clearly seen. Conversely, for $\Omega_0 = 0.08$ (panel (b1)), 
the spin components are almost equally populated, i.e. $N_\upar \approx N_\dar \approx 0.5$ and the quench produces only few percent oscillations. The quenching leads to change direction of the magnetic field, therefore it results in precession of the pseudo-spin component about the new axis. For the population-imbalanced case, the angle between the new field axis and spin is quite large, which leads to a change in sign of $\langle \sigma_z \rangle$ with time. On the other hand, for $\Omega_0 = 0.08$, the coupling dominates over the effective detuning, so the soliton is a combined superposition between two states, and the spin is nearly parallel to the new field after the quench. Therefore, spin-polarizations and populations remain almost constant with time.  

The IPR $\chi_j$ in the right panel of Fig.~\ref{fig:Nup-Nd-quench} represents the information about the spatial profile of each component behaves very differently in either cases. In (a2), the IPR starts decaying from the initial values. It is due to the fact that the self attraction scaled as $g_jN_j$, therefore the less populated component is weakly bound compared to the other component. So, the less populated component is mostly driven by the kinetic and the Rabi driven transfer from the majority component. The reduction of the IPR's demonstrates the large spin oscillation is accompanied by an increasing deformation of the soliton, which is the way the spin sector gives up energy to the spatial modes. For other case, the soliton behave rigidly, which is feebly effected by the quenching. As a result, it remain almost localized over time.

So far, we have discussed the Rabi quench-induced dynamics for a zero velocity soliton. Here, we demonstrate the equivalence of two realizations of quenching.  In Fig.~\ref{fig:spin-pol-Omega-quench}, we compare the time-dependent $\langle\sigma_i \rangle$, with $i = x,y, z$ of a soliton perturbing through a velocity in Eq.~\eqref{eqn:v_crit} with those of a soliton at rest after a sudden Rabi quench from $-\Omega_{0}$ to $\Omega_{0}.$ For $t <\tau$, the moving soliton is still in a region where the polarizations remain at the ground state values even though the soliton travels. The uniform motion of the soliton doesn't excite the spin dynamics because only the velocity perturbation without the spin-orbit coupling conserves the Galilean invariance. Upon arriving at the $x = 0$ barrier at $t = \tau$, the $\langle \sigma_x \rangle$ gradually goes to negative because the Rabi field changes sign just after $t > \tau$. At later time, the oscillation in $\langle\sigma_{i}\rangle$ correspond to spin precession around the new effective field axis. The agreement of all three components, with the correct phases and amplitudes and no adjustable parameter, shows that the velocity converts the spatial step into a temporal one, so the crossing acts as a ``dynamical quench''. Panel (b) shows the similar sequence of events on a time scale that is longer by a factor of two. Both $\Omega_0$ and $g_{\upar \upar}^2$ are halved, and so are all the energy scales while the ratio of coupling to detuning and the asymmetry in interactions remain same. The dynamics are therefore the rescaled version of those of (a): the first extrema of $\langle \sigma_j \rangle$ is found at $t - \tau \approx 50$ instead of $\approx 25$ in panel (a), with the same amplitude. In case of this dynamical quenching, the critical velocity can be defined following Eq. \eqref{eqn:v_crit}.  As an example, here we choose $v = 4v_{c}.$ Notice that the maximum change in $\langle\sigma_{z}\rangle\approx 0.3,$ corresponding well to the estimate 
for the maximum $\max(N_{\upar})\approx 576\Omega_{0}^{2}/g^{4},$ simultaneously shows that the exact linearization based on simple mapping of self-interaction onto $B\sigma_{z}$ field requires condition $\max(N_{\upar})\ll\,1,$ that is $\Omega_{0}/g^{2}\ll 0.03.$

\section{Conclusions and future outlook}

 \label{sec:4}

We have studied propagation of spin 1/2 soliton in a nonuniform barrier-like Rabi field with the Rabi coupling gradually changing from $-\Omega_{0}$ to $\Omega_{0},$ concentrating on the qualitative effects of the nonlinear interaction of soliton spin components. The spatial variation of the Rabi coupling at the barrier can be interpreted as an effective modulation of the local magnetic field, analogous to magnetic-field control in spintronic systems. 

Reflection and transmission of the soliton strongly depend on the symmetry of these nonlinear interactions and, in turn, on the soliton spin state. In the case of Manakov symmetry, where the nonlinearities depend on the total density only, the propagation is relatively simple both in terms of spin dynamics and velocity. For a lifted Manakov symmetry it becomes more complex with a considerable spin evolution, density redistribution between the components, strongly time-dependent soliton shape, and density radiation. 

The temporal evolution of the spin-projected inverse participation ratios further provides a quantitative characterization of the spatial evolution of the soliton profile.  We further find that the population-exchange rate is determined by the combined effect of the incident soliton velocity and the Rabi-coupling strength, providing means to control the internal spin dynamics. Notably, the critical velocity for soliton transmission is lower in the lifted Manakov-symmetry regime than in the Manakov-symmetric case, demonstrating the significant role of intercomponent interactions. Finally, we establish a connection between the finite-velocity soliton transmission dynamics and the well-known Rabi-quench evolution. This correspondence suggests that soliton motion through a spatially engineered Rabi barrier can provide an alternative route for accessing and controlling quench-like dynamics, thereby offering an additional mechanism for manipulating spin-resolved matter-wave transport.

The spatial variation of $\Omega(x)$ is similar to a domain wall in spintronics. The process is then analogous to injecting a spin from one layer of a spin valve into another with opposite magnetization. A mismatch between the spin and the new field produces a torque and the observed precession, with the velocity acting as a control parameter that selects the spin rotation. Since the precession starts after passing the barrier, the polarization at a given distance $L$ from it also depends on the velocity $v$ through the flight time $L/v,$ similar to a time-of-flight spin rotator. Therefore, the soliton acts as a nonlinear self-localized pseudospin carrier.

 Understanding the scattering process of spin-1/2 BEC-based solitons paves a venue to understanding and controlling of soliton-based spin states in nonuniform Rabi fields, including Rabi lattices, such as soliton binding and scattering and its relation to the symmetries of spin-related interactions.

 \section{Acknowledgments}
 S.K.S. would like to acknowledge the PARAM Shakti supercomputing facility provided at IIT Kharagpur, India.  E.Y.S. acknowledges support from PID 2021-126273NB-I00 and PID 2025-171880NB-I00 projects of MCIU/AEI/10.13039/501100011033 and ERDF “A way of making Europe” and the Basque Government through Grants No. IT1470-22 and IT1887-26.

\bibliography{references.bib}

\end{document}